\documentclass[aps,pre,reprint,superscriptaddress,showbibnotes,showpacs,10pt]{revtex4-1}
\usepackage{amsfonts}
\usepackage{amsmath}
\usepackage{amssymb}
\usepackage{graphicx}
\usepackage{color}
\begin{document}

\title{Stability and evolution of  wave packets in strongly coupled  degenerate plasmas}
\author{A. P. Misra}
\email{apmisra@visva-bharati.ac.in; apmisra@gmail.com}
\altaffiliation{Permanent address: Department of Mathematics, Siksha Bhavana, Visva-Bharati University, Santiniketan-731 235, India.}
\affiliation{Department of Physics, Ume\aa\ University, SE-901 87 Ume\aa, Sweden}
\author{P. K. Shukla}
\email{ps@tp4.rub.de; profshukla@yahoo.com}
\affiliation{RUB International Chair, International Center  for Advanced Studies in Physical Sciences, Faculty of Physics \& Astronomy, Ruhr University
Bochum, D-44780 Bochum, Germany}

\received{29 August 2011}
\revised{9 February 2012}
\begin{abstract}
We study the nonlinear propagation of electrostatic wave packets  in a collisional plasma composed of strongly coupled ions and relativistically degenerate electrons. The equilibrium of ions is maintained by an effective temperature associated with their strong coupling, whereas that of  electrons is provided by   the relativistic degeneracy pressure.  Using a multiple scale technique, a (3+1)-dimensional coupled set of nonlinear Schr\"{o}dinger-like equations with nonlocal nonlinearity is derived from a generalized viscoelastic hydrodynamic model.   These coupled equations, which govern the dynamics of  wave packets, are  used to study the oblique modulational instability of a Stoke's wave train to a small plane wave perturbation. We show that the wave packets, though  stable to the parallel modulation, becomes unstable against oblique  modulations. In contrast to the long-wavelength carrier modes,  the wave packets with short-wavelengths are shown to be stable  in the weakly relativistic case, whereas they can be stable or unstable in the  ultra-relativistic limit. Numerical simulation of the coupled equations reveals that a  steady state solution of the wave amplitude exists together with the  formation of a localized structure in (2+1) dimensions.  However, in the (3+1)-dimensional evolution,  a Gaussian   wave beam self-focuses after interaction and blows up in a finite time. The latter is, however,  arrested when the dispersion predominates over the nonlinearities. This  occurs when the Coulomb coupling strength is higher or a choice of obliqueness of modulation, or a wavelength of excitation is different.  Possible application of our results to the interior as well as in an outer mantle of white dwarfs are discussed.  
\end{abstract}


\maketitle

\section{Introduction}
The nonlinear propagation of electrostatic waves in strongly coupled plasmas has been of considerable interest in recent years because of its possible applications in compact astrophysical objects   (e.g. the interior of white dwarfs, neutron stars, the core of pre-Supernova stars), giant planetary interiors (e.g. Jupiter),   in laboratory  (e.g. ultra-cold plasmas by laser compression of matter)  as well as in nonideal plasmas for industrial applications (see e.g. some recent works \cite{solitary-UR2,sc-DA-collision,MI-sc2}).  {White dwarfs have typically masses approximately between $0.07$ and $8-10M_{\odot}$, where $M_{\odot}$  is the solar mass}. It has been predicted that the majority of white dwarfs have a core of carbon-oxygen composition, which itself is surrounded by a helium layer and, for most known white dwarfs, by an additional hydrogen layer. However, recent research has also indicated that there may be several white dwarfs with large volumes (or low density) primarily composed of carbon with little or no hydrogen or helium \cite{carbon-white-dwarf}.     {Furthermore, Koester in a recent study assumed that the core may be pure C or pure O \cite{C-O-white-dwarf}.} The core of these stars is, however, extremely dense  and consists of a plasma of unbounded nuclei and electrons, i.e.,    positively charged ions  providing almost all the mass (inertia)  and the pressure, as well as  electrons providing the pressure (restoring force) but none of the mass (inertialess). In these plasma environments, (typically at high densities $\sim 10^{34}$ cm$^{-3}$ and low temperatures $\sim 10^7$ K),  the ions form a regular lattice structure and electrons are relativistically degenerate with weak electrostatic interactions \cite{crystallization,compact-objects}. Nevertheless, in an outer mantle, plasmas may be composed of  non-relativistically degenerate electrons (at low-densities $\sim10^{26}$ cm$^{-3}$)  with strongly coupled ions.  In both these cases, the electron Fermi energy becomes larger compared to both the thermal energies $k_BT_{e,i}$ of electrons and ions as well as the typical electron-ion interaction energies. Therefore,  the  electron density, to a good approximation, can be assumed to be uniform and unaffected by the ions. 

On the other hand, above a certain mass ($M\approx1.4M_{\odot}$), the internal pressure of  white dwarfs becomes high enough for electrons to have sufficient momenta. In this case the relativistic effects become significant, and the  degenerate equation of state $P\sim n_e^{5/3}$ (weakly relativistic)  { that keeps the white dwarf from collapsing} under its own gravity,  changes to a different form $P\sim n_e^{4/3}$ (ultra-relativistic) \cite{compact-objects-Pressure}. In the latter case, the white dwarf collapses under self-gravitation  { into a denser object such as  a neutron star.} This change of degeneracy pressure (of course, there is no sharp transition) will certainly change the spectrum of collective oscillations of electrons and ions, which can be used to probe the density profiles and other characteristics of white dwarfs.   

Recent theoretical developments have indicated that strong correlations of ions significantly modify the dispersion properties of collective modes as well as the characteristics of nonlinear localized structures, e.g., solitons, shocks in strongly coupled degenerate plasmas \cite{solitary-UR2,sc-solitary-shock}. A number of works can be found in the literature dealing with the linear and nonlinear properties of wave modes in strongly coupled dusty plasmas \cite{sc-DA-collision,MI-sc2,sc-DA-modes1}. An addition, in a nonlinear regime, is the one dimensional propagation of wave packets and associated modulational instability \cite{mi-shukla} in strongly coupled dusty plasmas by Veeresha \textit{et al} \cite{MI-sc2}. Very recently, Shukla \textit{et al} \cite{solitary-UR2} have shown that the localization of ion modes  {is also possible in a one-dimensional strongly coupled} relativistically degenerate plasma.  It is, therefore, pertinent to investigate the influence of strongly coupled ions as well as the relativistic degeneracy pressure of electrons on the propagation of multi-dimensional wave packets in  collisional electron-ion plasmas,  {such as those  representative  of white dwarfs.} 

In contrast to (1+1)-dimensional evolution of envelope solitons, the nonlinear Schr\"{o}dinger equation in multi-dimensions [(2+1) or (3+1)] with cubic and/ or quadratic nonlocal nonlinearities may no longer be integrable. In this case, the system often exhibits wave collapse near the singularity even with a wide range of initial conditions instead of stable oscillations \cite{mi-ur}. A collapsing wave packet  with higher amplitudes thus self-focuses in shorter scales at the singularity   until other physical effects intervene to arrest it. Nevertheless, nonlocality has been known to  {give rise to novel phenomena} of generic natures.  For example, it may promote the modulational instability in self-defocusing media or can arrest wave collapse in multi-dimensional self-focusing media. Furthermore, nonlocal nonlinearity may affect   soliton interactions as well as may support the formation of stable localized structures \cite{solitons-multi}. It is thus of  interest to investigate the modulational instability  as well as the localization of  wave packets in strongly coupled  relativistically degenerate plasmas.
\section{Analysis}
\subsection{Basic equations}
We consider the nonlinear propagation of low-frequency ($kV_{Ti}<\omega<kV_{Te},\omega_{pi}$) electrostatic waves in an unmagnetized   collisional plasma consisting of inertial  {multi-charged} strongly coupled classical ions and inertialess  relativistically degenerate electrons with weak interparticle interactions. The dynamics of such waves can be described by a generalized hydrodynamic model which reads \cite{MI-sc2,sc-DA-modes1,solitary-UR2,review-shukla}
\begin{equation}
\frac{d n_{i}}{dt}+n_i\nabla\cdot\mathbf{v}_i=0,\label{be1}
\end{equation}
 \begin{eqnarray}
  &&\left(1+\tau_m\frac{d}{dt}\right)\left(m_i n_i\frac{d\mathbf{v}_i}{dt}+Z_ien_i\nabla\phi+\nabla P_i+m_in_i\nu_{in}\mathbf{v}_i\right) \notag\\
  && \hskip10pt =\eta\nabla^2\mathbf{v}_i+\left(\zeta+\frac{\eta}{3}\right)\nabla\left(\nabla\cdot\mathbf{v}_i \right),\label{be2}
   \end{eqnarray}
   \begin{equation}
   0=en_e\nabla\phi-\nabla P_e,\label{be3}
   \end{equation}
 \begin{equation}
 \nabla^{2}\phi=4\pi e\left(n_{e}-Z_in_{i}\right),\label{be4}
 \end{equation}
 where $n_j$, $\mathbf{v}_j$ and $m_j$ respectively denote the number density (with equilibrium value $n_{j0}$),  velocity and mass of different species ($j=e$ for electrons and $j=i$ for ions), $Z_i$ is the ion charge number, $e$ is the elementary charge and $\nu_{in}$ is the ion-neutral collision frequency.  Furthermore, $\phi$ is the electrostatic scalar potential, $P_e$ $(P_i)$ is the electron (ion) pressure to be defined later and $d/dt=\partial/\partial t+\mathbf{v}_i\cdot\nabla$ is the convective derivative. In equilibrium, the overall charge neutrality condition gives $n_{e0}=Z_in_{i0}$.  In Eq. \eqref{be2},   $\tau_m$ is the viscoelastic relaxation time given by \cite{sc-DA-modes1}
 \begin{equation}
 \tau_m=\frac{\zeta+4\eta/3}{n_{i0}k_BT_i}\left(1-\gamma_i\mu_i+\frac{4}{15}u(\Gamma_i)\right)^{-1}, \label{relaxation-time}
 \end{equation} 
where $T_i$ is the ion temperature, $k_B$ is the Boltzmann constant, $\zeta+4\eta/3$ is the coefficient of the effective ion viscosity in which  $\eta$ and $\zeta$ account respectively for the shear and bulk viscosity. Also,   $\gamma_i$ is the ion adiabatic index and $u(\Gamma_i)$ is a measure of the excess internal ion energy. { Here $\Gamma_i$, the ion coupling parameter (to be presented more extensively later in the text), is a measure of the ratio of the Coulomb energy to the kinetic energy per particle. For values of $\Gamma_i$ of the order of $1$ or larger, correlation effects become important and the plasma is then called strongly coupled.} The expression for $u(\Gamma_i)$ can be given for $\kappa\rightarrow0$  {(where $\kappa$, to be explained later, is the ratio of the interparticle distance to the ion Debye length)} as \cite{sc-u1,sc-u2}
\begin{equation}
 u\left(\Gamma_i\right)\approx\left\lbrace \begin{array}{lllllll}5
 -\frac{\sqrt{3}}{2}\Gamma_i^{3/2}  & \left(\Gamma_i<1\right), \\
 -0.90\Gamma_i+0.95\Gamma_i^{1/4}\\+0.18\Gamma_i^{-1/4}-0.80  & \left(1\leq\Gamma_i\leq160\right),\\
 1.5-0.90\Gamma_i+2980\Gamma_i^{-2} & \left(160<\Gamma_i\leq300\right),\\
 1.5-0.90\Gamma_i+9.6\Gamma^{-1}_i\\
 +840\Gamma_i^{-2}+1.1\times10^5\Gamma_i^{-3} & \left(300<\Gamma_i\leq2000\right).
 \end{array}      \right. \label{u-Gamma}
  \end{equation}
 The relaxation time $\tau_m$ represents a characteristic time scale to describe  {two classes} of wave modes with frequency $\omega$, namely the hydrodynamic modes with $\omega\tau_m\ll1$  and modes with $\omega\tau_m\gg1$, i.e.,  the kinetic modes. 
 The compressibility parameter  $\mu_i$ appearing in Eq. \eqref{relaxation-time} is    given by \cite{sc-DA-modes1}
\begin{equation}
\mu_i=\frac{1}{k_BT_i}\left(\frac{\partial P_i}{\partial n_i} \right)_{T_i}=1+\frac{1}{3}u(\Gamma_i)+\frac{\Gamma_i}{9}\frac{\partial u(\Gamma_i)}{\partial \Gamma_i}. \label{compressibility}
\end{equation}
Since $u(\Gamma_i)$ is negative for increasing values of $\Gamma_i$, $\mu_i$ can change its sign. It has been shown that for $\Gamma_i$ within the range $1<\Gamma_i<10$ this change of sign can cause the dispersion curve to turn over with the group velocity going to zero and then to negative values \cite{sc-DA-modes1}.  In the following subsections we will define few parameters that are relevant in the present theory.
\subsection{Degeneracy parameter}
The  degeneracy parameter for a particle of species $j$ can be defined as
\begin{equation}
\chi_{j}\equiv \frac{T_{Fj}}{T_{j}}=\frac{1}{2}\left(3\pi^2\right)^{2/3} \left(n_j\lambda_{Bj}^3\right)^{2/3}, \label{degeneracy}
\end{equation}
where $\lambda_{Bj}=\hbar/\sqrt{k_B T_j m_j}$ is the  thermal de Broglie wavelength, $E_{Fj}=k_B T_{Fj}=\hbar^2\left(3\pi^2 n_j\right)^{2/3}/2m_j$ is the Fermi energy and $T_{Fj}$ is the Fermi temperature. Thus, depending on the thermal energy $k_BT_j$, particles are said to be degenerate if the  number density $n_j$ exceeds the quantum concentration $n_{qj}\equiv\left(2m_jk_BT_j/\hbar^2\right)^{3/2}/3\pi^2$. Typically, for astrophysical dense plasmas, $n_e\gtrsim10^{27}$ cm$^{-3}$. So,  $\chi_e>1$ for $T_e\lesssim10^7$ K. However, in   metals ($n_e\sim10^{23}$ cm$^{-3}$)  electrons are degenerate at $T_e\lesssim10^5$ K. Thus, when the  {electrons form a degenerate system and ions are classical}, $\chi_e>1$ and $\chi_i<1$ must be satisfied. 
\subsection{Coupling parameter}
In order to measure the weak or strong interparticle interactions between electrons or ions we define two coupling parameters, namely the quantum coupling and the Coulomb coupling parameter. The quantum criterion of ideality  for degenerate electrons has the form
\begin{equation}
\Gamma_{e}=4\pi e^2 n_e^{1/3}/E_{Fe}\sim d/a_0, \label{quantum-coupling}
\end{equation}
where $d$  denotes the mean interparticle distance (Wigner-Seitz cell radius) given by $(4\pi/3)d^3n_j=1$ and $a_0$ is the Bohr radius. 
Since $E_{Fe}\propto n_e^{2/3}$, the parameter $\Gamma_e$ decreases  {below one} with increasing values of the electron density $(n_e\gtrsim10^{27}$ cm$^{-3})$. This implies that a degenerate electron plasma becomes even  more ideal with thermal compression. Moreover, at higher densities only electrons represent an ideal Fermi gas whereas the ion component is nonideal. Depending on the degree of its nonideality, one may look for ionic liquids, cellular or crystalline structures. Thus, one defines the criterion of ideality/ nonideality  by the ratio between the average potential energy of Coulomb interaction and the mean thermal energy characterized by the temperature $T_i$ for nondegenerate  {multi-charged} ions as
\begin{equation}
\Gamma\equiv\Gamma_ie^{-\kappa}=\left(\frac{Z_i^2e^2}{k_BT_id}\right)e^{-\kappa}, \label{Coulomb-coupling}
\end{equation} 
where the  factor $\kappa\equiv d/\lambda_{Di}$ measures the screening of the ion charge by the plasma over a distance of the ion Debye length $\lambda_{Di}=\sqrt{k_BT_i/4\pi n_{i0}Z_i^2e^2}$. Notice here that in a two-component electron-ion plasma in which electrons are degenerate and ions are classical, the screening distance of a  test charge is truly the ion Debye length, i.e.,
\begin{equation}
\lambda^{-2}_{scr}=\lambda^{-2}_{Fe}+\lambda^{-2}_{Di}\approx\lambda^{-2}_{Di}, \label{scr-length}
\end{equation}
where $\lambda_{Fe}=\left(k_B T_{Fe}/4\pi n_e e^2\right)^{1/2}$ is the Thomas-Fermi length (screening distance by degenerate electrons). Furthermore,  plasma represents a gaseous medium  for $\Gamma_i<1$, which changes to liquid-like $(\Gamma_i\gtrsim10)$  with density growth or with cooling. This plasma has a  short-range order. With further enhancement of $\Gamma_i$, the ion sub-system crystallizes in the range $(170\lesssim\Gamma_i\lesssim180)$, and for $\Gamma_i>180$, the plasma behaves like a solid with long-range order. In the latter, the Coulomb force completely dominates the dynamics, and  the ions arrange themselves in  a periodic lattice structure, which minimizes the electrostatic interaction energy. Typically, for a density $\sim10^{34}$ cm$^{-3}$ and a composition of pure O$^{16}$ in the  interior of a white dwarf, the resulting temperature for the onset of crystallization is $\sim10^7$ K  { \cite{crystal}}. 
\subsection{Degenerate equation of state}
For degenerate electrons, the energy distribution is no longer a thermal distribution, but one governed by the exclusion principle, i.e., the Fermi energy which has a pressure associated with it. The equation of state for relativistically degenerate electrons is given by \cite{compact-objects-Pressure} 
\begin{equation}
P_e=\frac{\pi m_e^4 c^5}{3h^3}\left[s\left(2s^2-3\right)\left(1+s^2\right)^{1/2}+3 \sinh^{-1}(s)\right], \label{pe}
\end{equation}
where $p_e=\left(3h^3n_e/8\pi\right)^{1/3}$ is the momentum of electrons on the Fermi surface, $h$ $(=2\pi\hbar)$ is the Planck's constant and $s=p_e/m_ec$ is the nondimensional parameter. Thus, in the weakly relativistic (or nonrelativistic)  limit ($s\ll1$) and ultra-relativistic limit ($s\gg1$) the pressure equation \eqref{pe}  reduces  to  two different forms:
\begin{equation}
P_e=\left\lbrace \begin{array}{lll}
\frac{1}{5}\frac{\hbar^2}{m_e}\left(3\pi^2\right)^{2/3}n_e^{5/3}=\frac{2}{5}E_{Fe}n_e & \text{for} & s\ll1,  \\
\frac{\hbar c}{4}\left(3\pi^2\right)^{1/3}n_e^{4/3} & \text{for} & s\gg1. 
\end{array} \right. \label{pe-wr-ur} 
\end{equation}
These pressures  can be combined to write
\begin{equation}
P_e=K_{\gamma}n_e^{\gamma}, \label{pe-comb}
\end{equation}
where 
\begin{equation}
K_{\gamma}=\frac{R_{\gamma}}{3\gamma}\left(3\pi^2\hbar^3\right)^{\gamma-1} \label{K-gamma}
\end{equation}
 with $R_{5/3}=1/m_e$ and $R_{4/3}=c$.  Thus, when the  Fermi energy becomes higher than the rest energy of electrons, the weakly relativistic pressure equation will no longer be valid, and one must use the ultra-relativistic equation of state.  {The latter in the case of massive white dwarf stars ($M\approx1.4M_{\text{sun}}$) gives rise to a lower degeneracy} pressure than the weakly relativistic case making them gravitationally unstable. 
 \subsection{Pressure equation for ions}
We now consider the pressure of ions as given by \cite{solitary-UR2}
\begin{equation}
\nabla P_i=\gamma_ik_B T_{if}\nabla n_i, \label{pressure-ion}
\end{equation}
where $T_{if}=T_{*}+\mu_i T_i$ is the effective ion temperature in which $T_{*}$ appears due to electrostatic interactions between  strongly coupled ions, and  is given by \cite{sc-pressure1,sc-pressure2} 
\begin{equation}
T_{*}=\frac{N_{nn}}{3}\Gamma_i T_i(1+\kappa)e^{-\kappa}. \label{temp-sc}
\end{equation}
Here,  $N_{nn}$ is determined by the ion structure and corresponds to the number of nearest neighbors (e.g., in the crystalline state, $N_{nn}=12$ for the fcc and hcp lattices, $N_{nn}=8$ for the bcc lattice).   Although, the parameter $\mu_i$ can be negative [\textit{cf.}  Eqs. (\ref{u-Gamma}) and (\ref{compressibility})] for increasing values of $\Gamma_i$,    $T_{*}$ may be comparable or even dominate over $\mu_iT_i$ for $\Gamma_i>>1$, and so the effective temperature $T_{if}$ $(>0)$ is most likely due to the strong coupling of ions. Typically, for $n_e=2\times10^{26}$ cm$^{-3}$, $T_e=40T_i=10^7$ K and $Z_i=8$ (relevant for weakly relativistic regime), we have $\Gamma_i\approx202$, $T_{*}=2\times10^8$ K, $\mu_iT_i=-2.4\times10^7$ K. Similarly, considering parameters in the ultra-relativistic regime, we find that $T_*$ is always a few orders of magnitude higher than the kinetic temperature of ions, $T_i$.
\subsection{Parameter regimes}
Our aim  is to consider a fully ionized two-component plasma in which electrons are relativistically degenerate with weak interaction (almost free) and  ions are multiply charged forming a classical system (nondegenerate) with strong electrostatic interaction. Since, $s\sim10^{-10}n_e^{1/3}$, in the weakly relativistic limit ($s\ll1$) we have  $n_e\lesssim10^{26}$ cm$^{-3}$. In this  regime, the degeneracy condition for electrons [\textit{cf.} Eq. \eqref{degeneracy}]  is   satisfied for $T_e\lesssim10^7$ K.  Also satisfied are the coupling parameters $\Gamma_e<1$, $\chi_i\ll1$, and $\Gamma_i\gg1$ for $Z_i>1$ and $T_i<T_e$. On the other hand, for ultra-relativistic degenerate electrons, we have $n_e\gtrsim10^{35}$ cm$^{-3}$ and $T_e\lesssim10^8$ K in order to satisfy $s\gg1$ and the degeneracy condition  \eqref{degeneracy}. In this case, ions are nondegenerate $(\chi_i<1)$ for $Z_i>1$ with a mass $m_i\sim Z_i m_0$, where $m_0=1.66\times 10^{-24}$ g is the  {unit of atomic mass, and the  conditions} $\chi_e\gg1$, $\Gamma_e\ll1$,   {$\Gamma_i\gtrsim100$} are satisfied. 

Again, it has been shown that   {the viscosity coefficient $C_{\text{vis}}\equiv(\zeta+4\eta/3)/n_{i0}k_BT_i$ in Eq. \eqref{relaxation-time} has a wide minima $(\sim1)$ in $1<\Gamma_i<10$, and it tends to increase as $\propto\Gamma_i^{4/3}$ for $\Gamma_i\geq10$. Also, it  becomes high for $\Gamma_i<1$ \cite{sc-viscosity}. Typical values of  $C_{\text{vis}}$ are  $\sim45$ for $\Gamma_i=0.1$,   $\sim46.4$ for $\Gamma_i=100$ and $\sim118.5$ for $\Gamma_i=202$. Thus, $\tau_m$ $(\propto C_{\text{vis}})$} becomes high in the weak coupling $(\Gamma_i<1)$  as well as in the high coupling $(\Gamma_i>1)$  regimes. So,  the kinetic modes exist only for $(\Gamma_i<1)$ or $(\Gamma_i\gg1)$ where the condition $\omega\tau_m\gg1$ is satisfied. In the range of $1<\Gamma_i<10$, $\tau_m$ is typically of the order of unity, and so the kinetic condition may no longer be valid. Again,   {since  $\Gamma_i>100$}  is more likely the case in both weakly relativistic and ultra-relativistic regimes  {[e.g. for parameters   relevant to the conditions  of  white dwarfs (C$^{12}$ and O$^{16}$ compositions) we have $\Gamma_i=125$ ($Z_i=6$) and $\Gamma_i=202$ ($Z_i=8$) for   $n_e=2\times10^{26}$ cm$^{-3}$, $T_e=40T_i=10^7$ K in the weakly relativistic case. In the ultra-relativistic regime, we have $\Gamma_i=165$ ($Z_i=6$) and $\Gamma_i=267$ ($Z_i=8$) for $n_e=10^{35}$ cm$^{-3}$, $T_e=2T_i=3\times10^8$ K]}, the hydrodynamic modes may not exist in the limit $\omega\tau_m\ll1$ by the same reason as for $\tau_m$. Thus, in our collisional plasmas we can specify the following two cases of interest:
 
Case I: $s\ll1$, $\chi_e>1$, $\chi_i<1$, $\Gamma_e<1$,   {$\Gamma_{i}>100$}  and $\omega\tau_m\gg1$. This represents the propagation of kinetic wave modes in a  plasma, composed of weakly relativistic degenerate electrons with weak interactions and strongly coupled classical ions. In this case, the valid regimes for the density and temperature to be relevant, e.g., in an outer mantle of a white dwarf are $n_e\lesssim10^{26}$ cm$^{-3}$, $T_i<T_e\lesssim10^7$ K with $Z_i>1$. 

Case II: $s\gg1$, $\chi_e>1$, $\chi_i<1$, $\Gamma_e<1$,   {$\Gamma_{i}>100$} and $\omega\tau_m\gg1$. That is, kinetic modes may exist in  a  plasma in which  degenerate electrons are ultra-relativistic  with weak interaction and ion components form strongly coupled classical system. The parameter regimes in this case, to be relevant in the core of a massive white dwarf are $n_e\gtrsim10^{35}$ cm$^{-3}$, $T_i<T_e\lesssim10^8$ K and $Z_i>1$.
 \subsection{Normalized system}
It is customary to normalize the physical quantities   by redefining them in terms of new variables: $N_j=n_j/n_{j0}$, $\mathbf{V}_i\equiv(U,V,W)=\mathbf{v}_i/V_{T}$, $T=t\omega_{pi}$ and $(X,Y,Z)=(x,y,z)/\lambda_{D}$, where $\lambda_D=\sqrt{\gamma_ik_BT_{if}/4\pi n_{i0}Z^2_ie^2}$ is the effective Debye length, $V_{T}=\sqrt{\gamma_ik_BT_{if}/m_i}$ is the  effective ion thermal speed and $\omega_{pi}=V_{T}/\lambda_{D}$ is the ion plasma frequency.  We then recast the basic equations \eqref{be1}, \eqref{be2} and \eqref{be4} in the following nondimensional forms. 
\begin{equation}
\frac{dN_{i}}{dT}+N_i\nabla\cdot\mathbf{V}_i=0,\label{ne1}
\end{equation}
 \begin{eqnarray}
  &&\left(1+\bar{\tau}_m\frac{d}{dT}\right)\left(N_i\frac{d\mathbf{V}_i}{dT}+D_{\gamma}N_i\nabla\phi_{\gamma}+\nabla N_i+\bar{\nu}_{in}N_i\mathbf{V}_i\right) \notag\\
  && \hskip10pt =\bar{\eta}\nabla^2\mathbf{V}_i+\left(\bar{\zeta}+\frac{\bar{\eta}}{3}\right)\nabla\left(\nabla\cdot\mathbf{V}_i \right),\label{ne2}
   \end{eqnarray}
    \begin{equation}
 \nabla^{2}\phi_{\gamma}=\left(N_{e}-N_{i}\right)/D_{\gamma}.\label{ne4}
 \end{equation}
Here $d/dT=\partial/\partial T+\mathbf{V}_i\cdot\nabla$ with $\nabla\equiv\left(\frac{\partial}{\partial X},\frac{\partial}{\partial Y},\frac{\partial}{\partial Z}\right)$,  $\bar{\nu}_{in}=\nu_{in}/\omega_{pi}$, $\bar{\tau}_m=\tau_m\omega_{pi}$,   $(\bar{\eta},\bar{\zeta})=(\eta,\zeta)\omega_{pi}/n_{i0}\gamma_ik_BT_{if}$, and the constant $D_{\gamma}$ is given for weakly relativistic $(\gamma=5/3)$ and ultra-relativistic $(\gamma=4/3)$ cases as:
\begin{equation}
 D_{\gamma}=\left\lbrace \begin{array}{lll}
 Z_iT_{Fe}/\gamma_iT_{if} & \text{for} & \gamma=5/3, \\
 \beta Z_im_e c^2/\gamma_ik_BT_{if} & \text{for} & \gamma=4/3,
 \end{array}      \right. \label{D-gamma}
  \end{equation}
where $\beta=\lambda_C\left(3\pi^2n_{e0}\right)^{1/3}$  is the dimensionless parameter with $\lambda_C=\hbar/m_ec$ denoting the reduced Compton wavelength. 
We are then left with Eq. (\ref{be3}), which can be integrated to obtain   the expression for the electron density  $N_e\equiv{n_e}/n_{e0}$ as
 \begin{equation}
 N_e=\left(1+\phi_{\gamma}\right)^{\frac{1}{\gamma-1}}\approx 1+A_{\gamma}\phi_{\gamma}+B_{\gamma}\phi^2_{\gamma}+C_{\gamma}\phi^3_{\gamma}, \label{Ne}
 \end{equation}
 where   $ \phi_{\gamma}$ $(<1)$ is different for different $\gamma$, and given by
 \begin{equation}
 \phi_{\gamma}=\left\lbrace \begin{array}{lll}
 e\phi/k_BT_{Fe} & \text{for} & \gamma=5/3, \\
 e\phi/\beta m_e c^2 & \text{for} & \gamma=4/3.
 \end{array}      \right. \label{phi-gamma}
 \end{equation}
 The coefficients appearing in Eq. \eqref{Ne} are
 \begin{eqnarray}
 &&A_{\gamma}=\frac{1}{\gamma-1}, \hskip10pt B_{\gamma}=\frac{2-\gamma}{2(\gamma-1)^2}, \notag\\ 
 && C_{\gamma}=\frac{(2-\gamma)(3-2\gamma)}{6(\gamma-1)^3}. \label{A-B-C}
 \end{eqnarray} 
\section{Perturbation method: Derivation of coupled equations} 
We consider the   propagation of   slowly varying weakly nonlinear wave envelopes propagating  with a group velocity $v_{g}$ oriented arbitrarily in the $xy$-plane to the direction of propagation.  Then in a coordinate
frame moving with the speed $v_{g}$, the space and the time variables can be stretched as \cite{mi-ur,MI-DS}
\begin{equation}
\xi=\epsilon(x-v_{gx}t),\eta=\epsilon (y-v_{gy}t), \zeta=\epsilon z, \tau=\epsilon^{2}t,\label{stretching}
\end{equation}
 where $\epsilon$ is a small parameter representing the strength of the wave amplitude, and $v_{gx}$, $v_{gy}$ are the components of the group velocity along the $x$ and $y$ axes. Since we are interested in the modulation of a plane wave as the carrier wave with the wave number $k$ and frequency  $\omega$, the dynamical variables can be expanded as
\begin{equation}
A=A_0+\sum_{n=1}^{\infty}\epsilon^{n}\sum_{l=-\infty}^{\infty}A_{l}^{(n)}(\xi,\eta,\zeta,\tau)e^{i(\textbf{k}\cdot\textbf{r}-\omega t)l},\label{expansion}
\end{equation}
where $A_0=1$ for $N_j$  and $A_0=0$ for other variables. Also, $A_{l}^{(n)}$  satisfies the reality condition $A_{-l}^{(n)}=A_{l}^{(n)\ast}$ with asterisk denoting the complex conjugate.  We then substitute the stretched variables from Eq. \eqref{stretching} and the expansion from Eq. \eqref{expansion} into the normalized equations \eqref{ne1}-\eqref{ne4}, and equate different powers of $\epsilon$ to obtain a set of reduced equations as given in the following subsections. Until and unless  mentioned we will omit the subscript $\gamma$ in $\phi_{\gamma}$ for simplicity, but we stress that the corresponding results will be different for different choice of $\gamma=5/3$ or $4/3$.
\subsection{Dispersion relation} 
 \begin{figure*}
\includegraphics[width=6in,height=4in,trim=0.0in 0in 0in 0in]{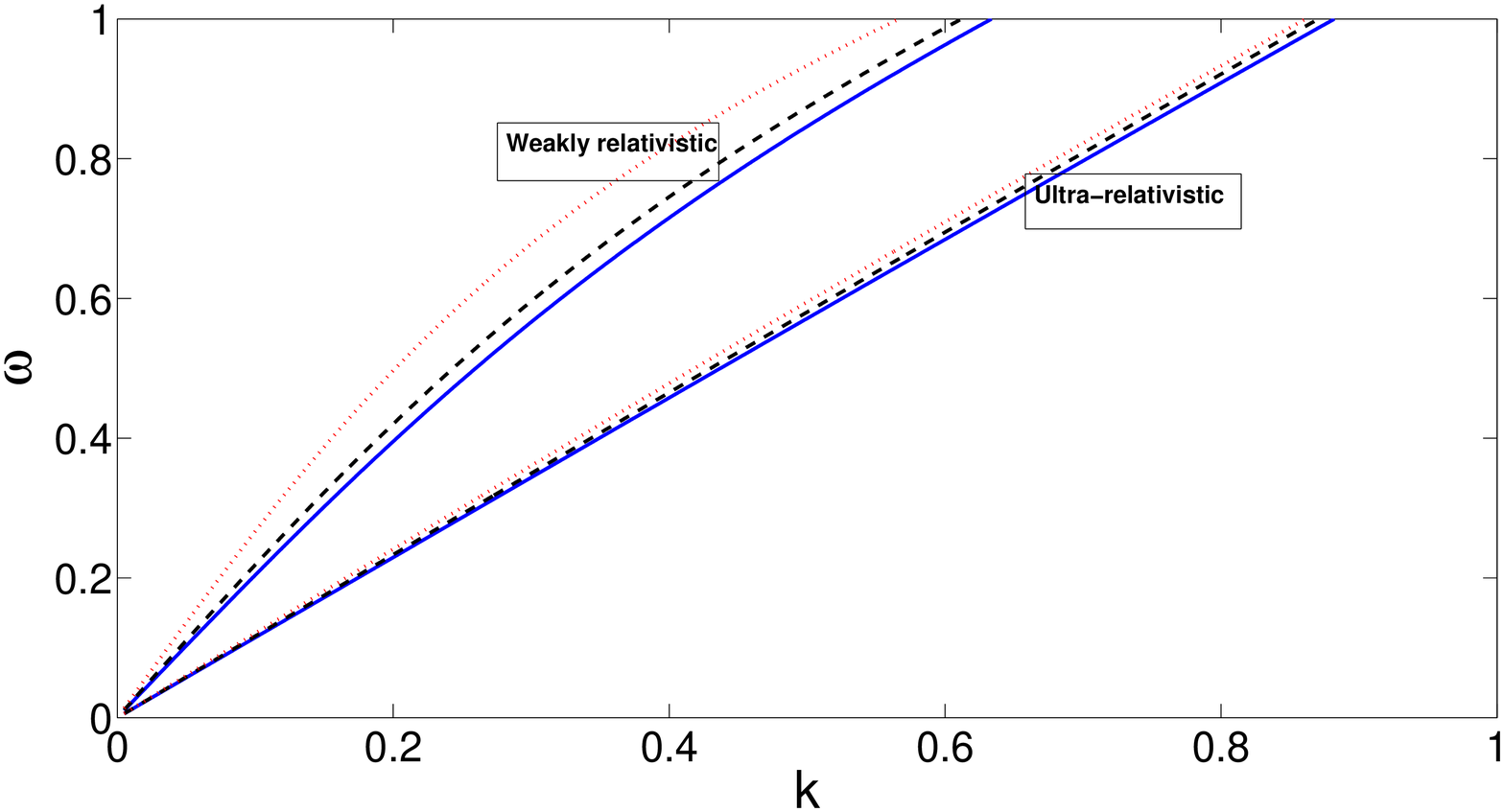}  
 \caption{ (Color online) Plot of the normalized wave frequency $\omega$ vs the normalized wave number $k$ [Eq. \eqref{dispersion-2}] for the kinetic modes.  The solid (blue), dashed (black) and dotted (red) lines respectively correspond to $\Gamma_i=125$, $202$, $512$ in weakly relativistic case, and  $\Gamma_i=165$, $267$, $678$ in ultra-relativistic regime as indicated in the figure. The other parameter values are as in Case I and Case II.  }
\end{figure*}
 In the lowest order of $\epsilon$ with $n=1$,   $l=1$, we obtain the following equations for the densities and velocities in terms of $\phi^{(1)}_{1}$.
\begin{equation}
N^{(1)}_{e1}=A_{\gamma}\phi^{(1)}_{1},\hskip5pt N^{(1)}_{i1}=K^2_{\gamma}\phi^{(1)}_{1}, \notag 
\end{equation}
\begin{equation}
\left(U^{(1)}_1,V^{(1)}_1\right)=\frac{\omega K^2_{\gamma}}{k^2}\left(k_x,k_y\right)\phi^{(1)}_{1},\hskip5pt W^{(1)}_{i1}=0, \label{1-order-eqs}
\end{equation}
where $K^2_{\gamma}=A_{\gamma}+D_{\gamma}k^2$. Since we are considering the modulation of a plane wave, $A_{l}^{(n)}$ are all set to zero except for $l=\pm1$. Thus, from Eq. \eqref{1-order-eqs} one readily obtains the following  dispersion relation:
\begin{equation}
1+\frac{1}{k^2\lambda^2_{p\gamma}}-\left({\omega^2-k^2+\frac{i\omega \eta^{*}k^2}{1-i\omega\bar{\tau}_m}}\right)^{-1}=0,\label{dispersion-law}
\end{equation} 
where $\lambda^2_{p\gamma}=A_{\gamma}/D_{\gamma}$ and $\eta^*=\bar{\zeta}+4\bar{\eta}/3$. The dispersion equation \eqref{dispersion-law} has the similar form as in Ref. \cite{sc-DA-modes1} for strongly coupled dusty plasmas.   {As said before, the   relaxation time $\tau_m$ defines two characteristic time scales to distinguish between two classes of wave modes, namely `hydrodynamic modes' ($\omega\tau_m\ll1$) and `kinetic modes' ($\omega\tau_m\gg1$). However as discussed in the previous section,  the hydrodynamic limit may not be satisfied for the parameter regimes   in the interior of white dwarfs.}     So, considering only the kinetic limit $\omega\tau_m\gg1$, Eq. \eqref{dispersion-law} reduces to
\begin{equation}
\omega^2=k^2\left(1+\frac{\eta^*}{\bar{\tau}_m}+\frac{\lambda^2_{p\gamma}}{1+k^2\lambda^2_{p\gamma}}\right). \label{dispersion-1}
\end{equation} 
Substitution of $\eta^*$ from Eq. \eqref{relaxation-time} into Eq. \eqref{dispersion-1} further gives
\begin{equation}
\omega=\pm k \left[\frac{T_*}{T_{if}}+\frac{T_i}{\gamma_iT_{if}}\left(1+\frac{4}{15}u\right)+\frac{\lambda^2_{p\gamma}}{1+k^2\lambda^2_{p\gamma}} \right]^{1/2}. \label{dispersion-2}
\end{equation}
For typical plasma parameters, as in Case I or Case II, the first term $(\sim1)$ in Eq. \eqref{dispersion-2} under the square root is larger than the second term $(\sim-0.03)$, i.e., the contribution of $T_*$ is always greater than that from $T_i$, and  hence   $\omega$ is always real for all $k$.  However, in absence of $T_*$, and since the second term may be  negative,   there must exist a critical wave number $k_c$ below which the kinetic modes exist.   
Furthermore,   {Eq. \eqref{dispersion-2}  shows that the phase velocity of the carrier modes}, $\omega/k>1$ or,  $\omega>kV_T$ (in dimensional form). Again, since the term in the square root is typically $\gtrsim1$ for $\Gamma_i\gg1$,  the low-frequency $(<\omega_{pi})$ wave propagation is possible for $k<1$. Thus, in both the weakly and ultra-relativistic regimes as in Case I and II, $\omega$ approaches unity for $k<1$. The latter also ensures that the wavelength is greater than the effective Debye length for the collective behaviors of the plasma not to disappear.  Equation \eqref{dispersion-2} also clears that the wave frequency increases with $k$, and it approaches the plasma frequency  $\omega_{pi}$ (i.e. $\omega\approx1$) at $k\approx0.5$. For the behaviors of the modes we plot $\omega$ vs $k$ as in Fig. 1. In both the weakly relativistic and ultra-relativistic cases, the wave frequency  {is shown to increase with increasing  coupling parameter} $\Gamma_i$.

  {We note that in the kinetic regime, the ion modes do not experience any viscous damping. The latter can, however, occur in the hydrodynamic regime. Furthermore, the ion coupling corrections appear through $u\left(\Gamma_i\right)$ whose estimates can be obtained from Eq. \eqref{u-Gamma}. Since $u\left(\Gamma_i\right)$ can be negative for increasing values of $\Gamma_i$, these corrections can lead to the turnover effect in which the frequency (and hence the group velocity) goes to zero and then to negative values \cite{sc-DA-modes1}. However, this is not the case for the kinetic wave modes as evident from Fig. 1. Such   effects can be strong in the `hydrodynamic limit' \cite{sc-DA-modes1}. The term $\propto \lambda_{p\gamma}$ appears due to degeneracy pressure of electrons in weakly relativistic $(\gamma=5/3)$ and ultra-relativistic $(\gamma=4/3)$ limits. As is clear from Fig. 1 that the influence of this term on the dispersive curves is, however, weaker in the ultra-relativistic limit than that in  the weakly relativistic one.   There is an additional correction term ($\propto T_*$) which arises due to electrostatic interaction of strongly coupled ions. We further note that the kinetic modes exist only in the strong coupling regime ($\Gamma_i>100$, i.e. close to crystallization) where the condition $\omega\tau_m\gg1$ is satisfied and $\tau_m$ can be quite large. Physically, the existence of large values of $\tau_m$ implies strong memory effects in the medium and hence the predominance of elastic effects \cite{MI-sc2}. In contrast to the weakly relativistic case,  the dispersion curves (see Fig. 1) of the ion modes for different values of $\Gamma_i$ show that the wave frequency $\omega$ typically varies linearly with $k<1$ in the ultra-relativistic limit. We also find that the effect of  the ion coupling parameter $\Gamma_i$ on the kinetic modes is, however,   stronger in the weakly relativistic limit than the ultra-relativistic case.} 
\subsection{Group velocity}
 \begin{figure*}
\includegraphics[width=6in,height=4in,trim=0.0in 0in 0in 0in]{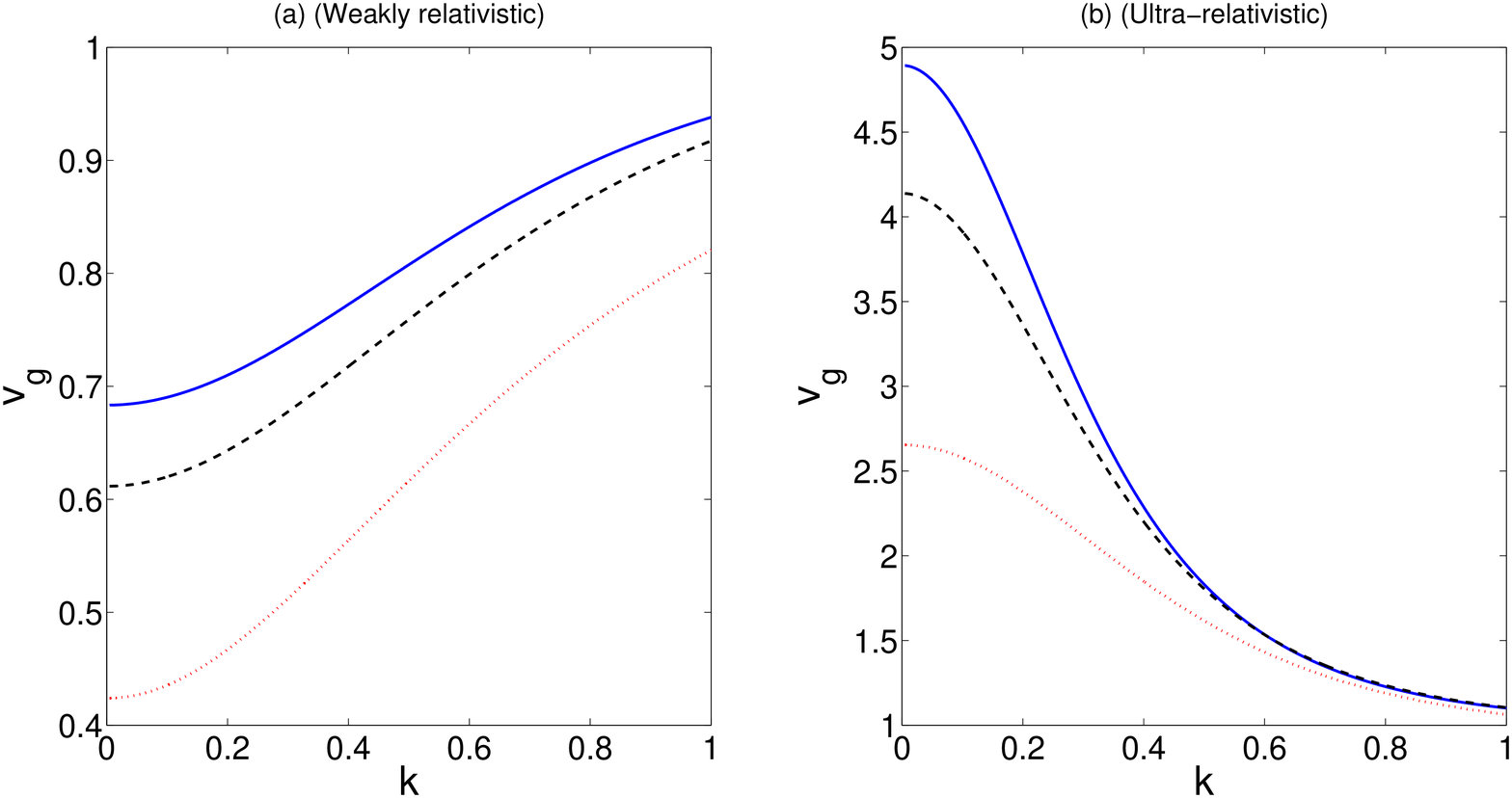}  
 \caption{ (Color online) The normalized group velocity dispersion $v_g$ [Eq. \eqref{group-velocity-2}] is plotted against the normalized wave number $k$  for the kinetic modes.  The solid (blue), dashed (black) and dotted (red) lines are for different values of $\Gamma_i$ as in Fig. 1. }
\end{figure*}
In the second order expressions for $n=2$, $l=1$ we obtain the corrections for $N^{(2)}_{e1}$, $N^{(2)}_{i1}$ etc.  in terms of $\phi^{(2)}_1$ and $\phi^{(1)}_1$.   {After eliminating those $N^{(2)}_{e1}$, $N^{(2)}_{i1}$ etc., we obtain a resulting equation in which  the coefficient of $\phi^{(2)}_1$   vanishes by the dispersion relation}, and  the coefficients of $\partial \phi^{(1)}_1/\partial\xi$ and $\partial \phi^{(1)}_1/\partial\eta$ are set to zero to obtain the group velocity components (compatibility condition) as
\begin{equation}
\frac{\partial\omega}{\partial k_{x,y}}\equiv{v_{g(x,y)}}=\frac{2k_{x,y}}{\tilde{\omega}}\left(1-\frac{i\omega\eta^{*}}{1-i\omega\bar{\tau}_m}+\frac{A_{\gamma}D_{\gamma}}{K^4_{\gamma}} \right), \label{group-velocity-1} 
\end{equation} 
where
\begin{equation}
\tilde{\omega}=2\omega+ \frac{i\eta^{*}k^2}{(1-i\omega\bar{\tau}_m)^2}. 
\end{equation}
The expression for  the group velocity can further be simplified in the limit of $\omega\tau_m\gg1$ and using Eq. \eqref{relaxation-time}. For $\omega\tau_m\gg1$, we have $\tilde{\omega}=(2k^2/\omega)\left[\omega^2/k^2-i(\eta^*/\bar{\tau}_m)/2\omega \bar{\tau}_m\right]$.  Since $\eta^*/\bar{\tau}_m\ll1$ for $\Gamma_i\gg1$; also $\omega\tau_m\gg1$ (already assumed) and $\omega/k>1$ from the dispersion relation, the imaginary part of $\tilde{\omega}$ can be neglected compared to the real part. Thus, we obtain
\begin{equation}
v_{g}\approx \frac{k}{\omega}\left[\frac{T_*}{T_{if}}+\frac{T_i}{\gamma_iT_{if}}\left(1+\frac{4}{15}u\right)+\frac{A_{\gamma}D_{\gamma}}{K^4_{\gamma}} \right]. \label{group-velocity-2}
\end{equation}
From Fig. 2 we find that in the weakly relativistic limit [Fig. 2(a)], the modes with short-wavelengths have greater group velocity than those with long-wavelengths, and this group velocity is always smaller than the effective ion thermal speed as well as the phase speed of the carrier wave.   {Also, in this case $v_g$ increases as the wave number $k$ $(<1)$ increases. However, as $k$ approaches $0$ (i.e. in the long wavelength limit), $v_g$ tends to a steady state value. This anomalous group velocity dispersion is an important condition for the modulation instability of wave packets. In contrast to the weakly relativistic case,    the wave modes with long-wavelengths propagate with higher group velocity than those with short-wavelengths in the ultra-relativistic limit [Fig. 2(b)].} Also, the group velocity in this case is greater  than the ion thermal speed and the phase speed except for some higher values of $\Gamma_i\sim10^4$. The latter may not be relevant in  in the interior of white dwarfs as values of $Z_i$ are not so large there.   {Furthermore, in the ultra-relativistic limit, the group velocity tends to decrease with increasing values of $k<1$, and there exists a critical value $k_c<1$ of $k$ above which the behaviors of $v_g$ remain almost the same even with different values of the ion coupling parameter $\Gamma_i$.}   In both these weakly relativistic and ultra-relativistic cases,  {the higher the coupling parameter $\Gamma_i$, the lower} is the group velocity of dispersion.  

\subsection{Coupled equations}
Considering the zeroth harmonic modes for $n=2,3$, $l=0$, which appear due to the nonlinear self-interaction of the carrier waves, we obtain corrections  for the densities and velocities (see for details, Appendix-B). These expressions are then used to eliminate  the variables to obtain the following equation for $\phi^{(2)}_0$.
\begin{widetext}
\begin{equation}
\left(R_1\frac{\partial^2}{\partial\xi^2}+R_2\frac{\partial^2}{\partial\eta^2}+R_3\frac{\partial^2}{\partial\zeta^2}+R_4\frac{\partial^2}{\partial\xi\partial\eta}\right)\phi^{(2)}_0=\left(S_1\frac{\partial^2}{\partial\xi^2}+S_2\frac{\partial^2}{\partial\eta^2}+S_3\frac{\partial^2}{\partial\zeta^2}+S_4\frac{\partial^2}{\partial\xi\partial\eta}\right)\lvert\phi^{(1)}_1\rvert^2, \label{nonlocal}
\end{equation}
where the coefficients are given by
\begin{equation}
R_{1,2}=A_{\gamma}\left(v^2_{g(x,y)}-1\right)-D_{\gamma},\hskip5pt R_3=-(A_{\gamma}+D_{\gamma}),\hskip5pt R_4=2A_{\gamma}v_{gx}v_{gy},
\end{equation}
\begin{equation}
S_{1,2}=2B_{\gamma}\left(1-v^2_{g(x,y)}\right)+\frac{\omega K^4_{\gamma}}{k^2}k_{x,y}\left(v_{g(x,y)}-\frac{\omega k_{x,y}}{k^2}\right)+D_{\gamma}K^2_{\gamma},
\end{equation}
\begin{equation}
S_3=2B_{\gamma}+D_{\gamma}K^2_{\gamma}, \hskip5pt S_4=\frac{\omega K^4_{\gamma}}{k^2}\left(k_xv_{gy}+k_yv_{gx}+2\omega\frac{k_xk_y}{k^2}\right)-4B_{\gamma}v_{gx}v_{gy}.
\end{equation}
\end{widetext}
 {We ought to mention that in the coefficient of $\epsilon^2$ for $n=2,l=0$, one usually obtains from the momentum balance equation for ions the relation}  $\bar{\nu}_{in}\lvert \phi^{(1)}_1\rvert^2=0$. Since $\lvert \phi^{(1)}_1\rvert^2$ is of the order of $\epsilon^2$, $\bar{\nu}_{in}$  should be at least of the order of $\epsilon$, and so it will contribute to the coefficient of $\epsilon^3$ of the momentum equation for $n=3,l=1$. 

Next, the second order harmonic modes for $n=2,l=2$ are obtained as
\begin{equation}
U^{(2)}_2=A_{22}[\phi^{(1)}_1]^2, \hskip5pt V^{(2)}_2=B_{22}[\phi^{(1)}_1]^2,\hskip5pt W^{(2)}_2=0,
\end{equation}
\begin{equation}
N^{(2)}_{e2}=A_{\gamma}\phi^{(2)}_2+B_{\gamma}[\phi^{(1)}_1]^2, 
\end{equation}
\begin{equation}
N^{(2)}_{i2}=N^{(2)}_{e2}+4k^2D_{\gamma}\phi^{(2)}_2, \hskip5pt\phi^{(2)}_{2}=D_{22}[\phi^{(1)}_1]^2,
\end{equation}
where the coefficients are given in Appendix C. Finally, for  $n=3,l=1$ we obtain expressions for the third-order first harmonic modes. Here the coefficients of $\phi^{(3)}_1$ and $\phi^{(2)}_1$ vanish by the dispersion relation. Thus,  {we obtain after a few steps} the following nonlocal nonlinear Schr\"{o}dinger equation:
\begin{eqnarray}
 &&i\frac{\partial\phi^{(1)}_1}{\partial\tau}+P_{1}\frac{\partial^{2}\phi^{(1)}_1}{\partial\xi^{2}}+P_{2}\frac{\partial^{2}\phi^{(1)}_1}{\partial\eta^{2}}+P_{3}\frac{\partial^{2}\phi^{(1)}_1}{\partial\zeta^{2}}+P_4\frac{\partial^{2}\phi^{(1)}_1}{\partial\xi\partial\eta}\notag \\
 && +Q_{1}|\phi^{(1)}_1|^{2}\phi^{(1)}_1+Q_{2}\phi^{(2)}_0\phi^{(1)}_1+iQ_3\phi^{(1)}_1=0\label{NLSE}.
\end{eqnarray}

In deriving Eq. \eqref{NLSE}, we have neglected a small contribution from the term $\left(k_xU^{(2)}_0+k_yV^{(2)}_0\right)\phi^{(1)}_1$ compared to ${\partial\phi^{(1)}_1}/{\partial\tau}$. Because, from the dispersion relation we find that the phase speed of the wave is larger than the ion thermal speed. Also, the modes $U^{(2)}_0$ and $V^{(2)}_0$, which arise due to the mean motion of ions, can even be smaller than the ion thermal speed, since they appear as small (higher order of $\epsilon$) corrections. So, $\omega\gg k_xU^{(2)}_0+k_yV^{(2)}_0$ can be a good approximation to the present case. 

The terms with coefficients $P_1$, $P_2$ appear due to wave group dispersion, those with $P_3$ and $P_4$ are respectively due to the departure from the two-dimensional formalism and due to the oblique modulation.  The cubic nonlinearity (Kerr) is due to the carrier wave self-interaction originating from the zeroth harmonic modes (or slow modes), and the   nonlocal nonlinear  (quadratic) one appears due to the coupling between the dynamical field associated with the first harmonic (with a  {`cascade effect'} from the second harmonic) and a static field generated due to the mean motion (zeroth harmonic) in plasmas.  

Now, Eq. \eqref{NLSE} is coupled to Eq. \eqref{nonlocal}, and  we recast the two equations as [considering now that $\phi^{(1)}_1$, $\phi^{(2)}_0$ are dependent on  $\gamma$, i.e. rewriting  $\Phi_{\gamma}\equiv\phi^{(1)}_1$ and $\Psi_{\gamma}\equiv\phi^{(2)}_0$ for weakly relativistic $(\gamma=5/3)$ and ultra-relativistic $(\gamma=4/3)$ cases] follows:
\begin{widetext}
\begin{equation}
i\frac{\partial\Phi_{\gamma}}{\partial\tau}+P_{1}\frac{\partial^{2}\Phi_{\gamma}}{\partial\xi^{2}}+P_{2}\frac{\partial^{2}\Phi_{\gamma}}{\partial\eta^{2}}+P_{3}\frac{\partial^{2}\Phi_{\gamma}}{\partial\zeta^{2}}+P_4\frac{\partial^{2}\Phi_{\gamma}}{\partial\xi\partial\eta}
+Q_{1}|\Phi_{\gamma}|^{2}\Phi_{\gamma}+Q_{2}\Psi_{\gamma}\Phi_{\gamma}+iQ_3\Phi_{\gamma}=0\label{NLSE-1},
\end{equation}
\begin{equation}
R_1\frac{\partial^2\Psi_{\gamma}}{\partial\xi^2}+R_2\frac{\partial^2\Psi_{\gamma}}{\partial\eta^2}+R_3\frac{\partial^2\Psi_{\gamma}}{\partial\zeta^2}+R_4\frac{\partial^2\Psi_{\gamma}}{\partial\xi\partial\eta}=S_1\frac{\partial^2\Phi_{\gamma}}{\partial\xi^2}+S_2\frac{\partial^2\Phi_{\gamma}}{\partial\eta^2}+S_3\frac{\partial^2\Phi_{\gamma}}{\partial\zeta^2}+S_4\frac{\partial^2\Phi_{\gamma}}{\partial\xi\partial\eta}, \label{NLSE-2}
\end{equation}
\end{widetext}
in which    {the coefficients are different for different values of $\gamma=5/3,4/3$ (according as the degenerate electrons are weakly relativistic or ultra-relativistic)} and are given in Appendix D.
\section{Modulational instability}
 \begin{figure*}
\includegraphics[width=6in,height=4in,trim=0.0in 0in 0in 0in]{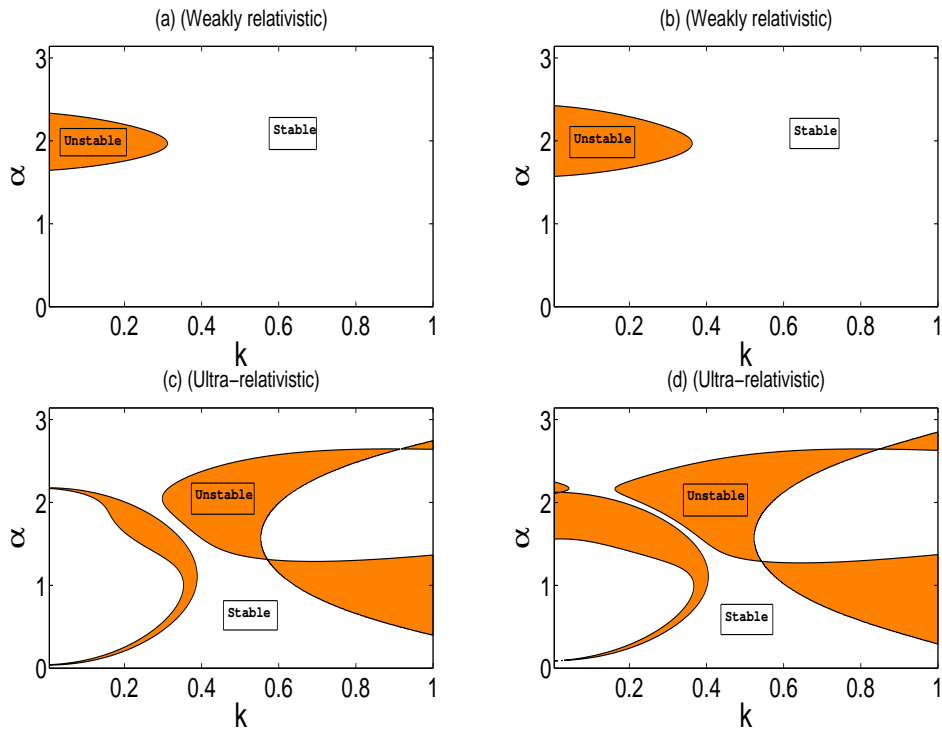}  
 \caption{ (Color online) The stable (blank or white) and unstable (shaded or colored) regions are shown in the plane of normalized wave number $k$ and the obliqueness of modulation $\alpha$, corresponding to the conditions $\Upsilon<0$ and $\Upsilon>0$ respectively as in the text.   Subplots (a), (b) are for $\Gamma_i=125$, $202$ (weakly relativistic), and (c), (d) (Ultra-relativistic)  are for  $\Gamma_i=165$, $267$ respectively. Other parameter values are as in Case I and Case II, and $K=0.001$, $\alpha_1=2$, $\alpha_2=0.1$,  $\phi_0=0.01$.}
\end{figure*}
\begin{figure*}
\includegraphics[width=6in,height=4in,trim=0.0in 0in 0in 0in]{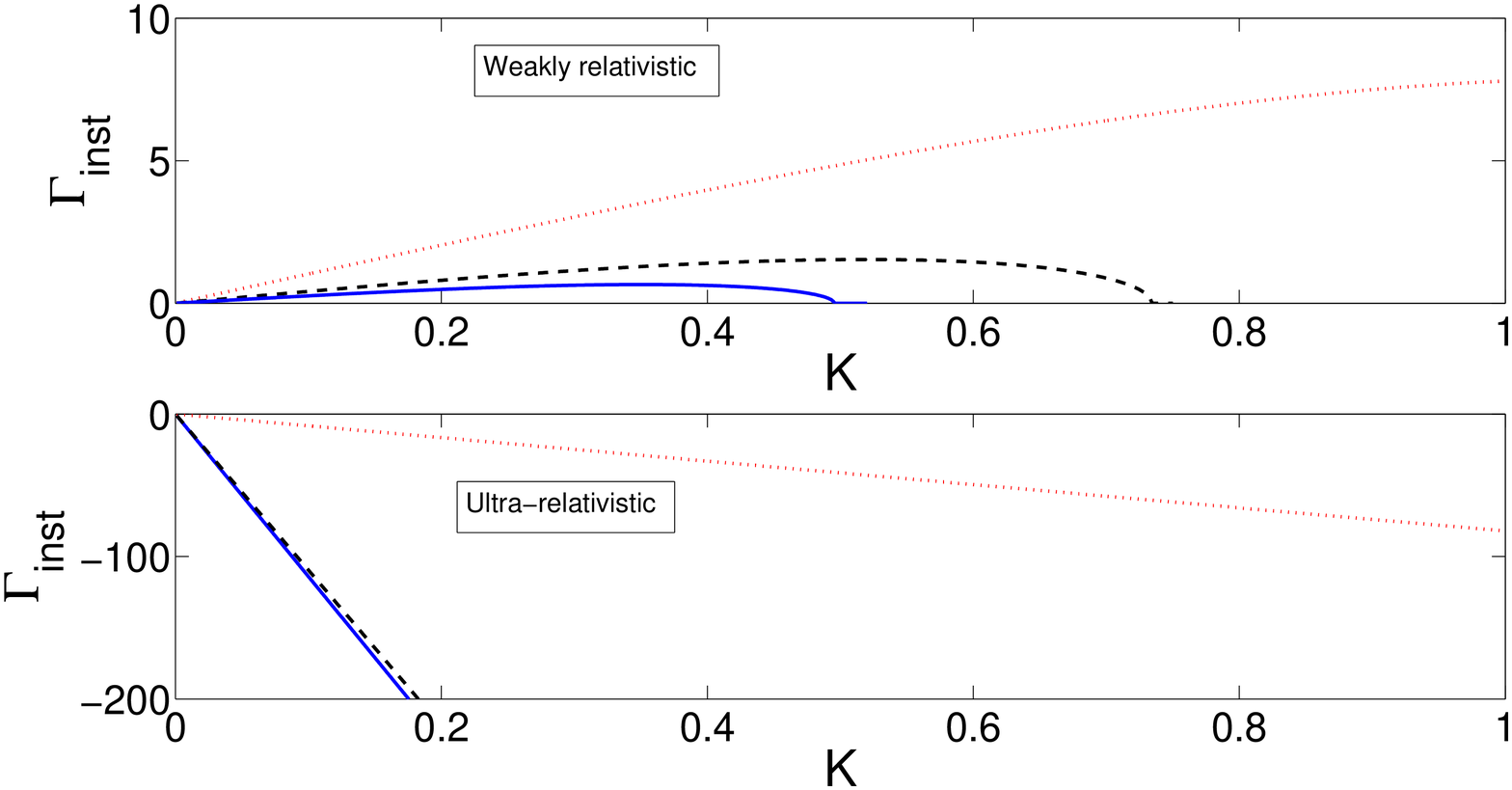}  
 \caption{ (Color online) The growth (case of weakly relativistic, see upper panel) and decay (case of ultra-relativistic, see lower panel) rates [Eq. \eqref{growth-rate}] are shown at $\alpha=2$ and $\alpha=1.7$ respectively. The solid (blue), dashed (black) and dotted (red) lines are for different values of $\Gamma_i$ as in Fig. 1. The other parameters are as in Case I and Case II, and $k=0.2$, $\alpha_1=2$, $\alpha_2=0.1$, $\phi_0=0.01$.}
\end{figure*}
  {The modulation of slowly varying wave amplitudes may occur due to, e.g., parametric wave coupling, nonlinear interaction between high- and low-frequency modes or, self-interaction of the carrier wave modes. These phenomena, however, have relevance with the MI, which constitutes one of the most fundamental effects associated with the wave propagation in nonlinear media.  Such instability basically signifies the exponential growth or decay of a small perturbation of the wave  as it propagates in plasmas. The gain leads to amplification of sidebands, which break up the otherwise uniform wave and lead to energy localization via the formation of localized structures. Thus, the MI may act as a precursor for the formation of bright envelope solitons. On the other hand, the formation of dark solitons requires the absence of MI in the constant intensity background.  The MI can, however, be affected by the obliqueness of modulation, the electron degeneracy as well as the strong coupling effects of ions which we will discuss shortly. } 

Let us consider the modulation of  a plane wave solution of Eqs. \eqref{NLSE-1} and \eqref{NLSE-2} of the form (omitting the subscript $\gamma$ once again, for simplicity)
$\Phi=\Phi_{0} e^{-i\Omega_0\tau}$, $\Psi=0$  so that $\Omega_0=-Q_{1}\Phi_{0}^{2}-iQ_{3}$, where $\Phi_0$ is a constant. Here the choice of $\Psi$ is immaterial as the stability condition does not depend on it, and the solution for $\Phi$ is not unique.  We then modulate the wave amplitude as a plane wave perturbation with frequency $\Omega$ and wave number $K$, i.e., 
 $\Phi=\left(\Phi_{0}+\Phi_1e^{i\mathbf{K}\cdot\mathbf{R}-i\Omega\tau}+\Phi_2e^{-i\mathbf{K}\cdot\mathbf{R}+i\Omega\tau}\right)e^{-i\Omega_0\tau}$ and       
$\Psi=\Psi_1e^{i\mathbf{K}\cdot\mathbf{R}-i\Omega\tau}+\Psi_2e^{-i\mathbf{K}\cdot\mathbf{R}+i\Omega\tau}$, where $\Phi_{1,2}$, $\Psi_{1,2}$ are all real constants,   
 $\mathbf{K}\equiv (K_1,K_2,K_3)$   and $\mathbf{R}\equiv(\xi,\eta,\zeta)$. 
Looking for the nonzero solution of the small
amplitude  perturbations, we  obtain from Eqs. (\ref{NLSE-1}) and (\ref{NLSE-2}) the following dispersion relation for the modulated  {wave packet}
 \begin{equation}
\Omega^{2}=f_1^2\left[1-\frac{2\Phi^2_0}{f_1}\left(Q_1+\frac{Q_2f_3}{f_2} \right)\right],\label{dispersion-modulation-1}
\end{equation}
 where $f_1$, $f_2$ and $f_3$ are given by 
 \begin{eqnarray}
 &&f_{1,2,3}=(P_1,R_1,S_1)K_1^2+(P_2,R_2,S_2)K_2^2 \notag \\
 &&+(P_3,R_3,S_3)K_3^2+(P_4,R_4,S_4)K_1K_2. 
 \end{eqnarray}
 In the limit of $\omega\tau_m\gg1$, all the coefficients in Eqs. \eqref{NLSE-1} and \eqref {NLSE-2} are real and so, Eq. \eqref{dispersion-modulation-1} can be rewritten as 
  \begin{equation}
\Omega^{2}=f_1^2\left(1-\frac{K^2_c}{K^2}\right),\label{dispersion-modulation-2}
\end{equation}
 where  $K_c$ is the critical wave number given by
  \begin{equation}
K_{c}^{2}=\frac{2\Phi_{0}^{2}}{\tilde{f}_1\tilde{f}_2}\left(Q_{1}\tilde{f}_2+Q_{2}\tilde{f}_3\right)\left(1+\alpha_1^{2}+\alpha^2_2\right).\label{critical-wave-number}
\end{equation}
Here, $\tilde{f}_{1,2,3}=f_{1,2,3}/K^2_1$ and  $\alpha_{1,2}=K_{2,3}/K_{1}$ define the obliqueness of perturbation of the wave numbers.  {As we will see later, those perturbations} can change the stable and unstable regions for wave packets.  Equation (\ref{dispersion-modulation-2}) shows that the MI
sets in for  $K<K_{c}$, and the right-hand
side of Eq. (\ref{critical-wave-number}) is positive, i.e. $\Upsilon\equiv \left(Q_{1}\tilde{f}_2+Q_{2}\tilde{f}_3\right)/\tilde{f}_1\tilde{f}_2>0$.  {In this case the perturbations grow or decay exponentially during propagation of waves.} On the other hand,   
for $K>K_c$, the wave packet is said to be  stable under the modulation. The instability growth or decay rate (letting
$\Omega=i\Gamma_{\text{inst}})$ can be obtained as
 \begin{equation}
\Gamma_{\text{inst}}=\frac{K^{2}\tilde{f}_1}{1+\alpha_1^{2}+\alpha^2_2}\sqrt{\frac{K_{c}^{2}}{K^{2}}-1},\label{growth-rate}
\end{equation}
where the maximum value is achieved at $K=K_{c}/\sqrt{2}$, and  is  given by 
\begin{equation}
\Gamma^{(\text{max})}_{\text{inst}}=\frac{\Phi_{0}^{2}}{\tilde{f}_2}\left(Q_{1}\tilde{f}_2+Q_{2}\tilde{f}_3\right).
\end{equation}
We numerically investigate the stable and unstable regions for  {the plane wave solution relying on the above condition} and with different coupling parameters $\Gamma_i$ as applicable for weakly relativistic as well as  ultra-relativistic regimes (\textit{cf.} Case I and Case II). We consider the parameter values which are relevant to the conditions  of  white dwarfs. For example, in the weakly relativistic case, $n_e=2\times10^{26}$ cm$^{-3}$, $T_e=40T_i=10^7$ K and vary $Z_i=6$ ($\Gamma_i=125$), $8$ ($\Gamma_i=202$). In the ultra-relativistic case, we take $n_e=10^{35}$ cm$^{-3}$, $T_e=2T_i=3\times10^8$ K with $Z_i=6$ ($\Gamma_i=165$) and $8$ ($\Gamma_i=267$) for C$^{12}$ and O$^{16}$ compositions. Due to highly efficient thermal conductivity properties of the degenerate electron gas, the interior of a white dwarf is nearly isothermal with a temperature $\sim10^7$ or $10^8$ K. 

Since $K$ represents the wave number of perturbation it should not be higher than the carrier wave number $k$. However, $\alpha_1$ and $\alpha_2$ can even be larger than  unity, and thereby can change the sign of $\Upsilon$, and hence the stable and unstable regions. For example, for a fixed $\Gamma_i=125$ as in Fig. 3(a), the wave is  stable  in the regimes, namely  $0\leq\alpha_1\lesssim0.5$ and $\alpha_2\geq0$; $0\leq\alpha_1\lesssim0.6$ and $\alpha_2\geq0.2$; $0\leq\alpha_1\lesssim0.7$ and $\alpha_2\geq0.4$  etc. In the other regimes of $\alpha_1$ and $\alpha_2$, the wave can be unstable. Fig. 3 shows different stable (white or blank) and unstable (colored or shaded) regions in which panels (a), (b) are for the weakly relativistic case with different $\Gamma_i=125$ and $202$ respectively, and panels (c), (d) for $\Gamma_i=165$ and $267$ in the ultra-relativistic regime.  {From Figs. 3(a) and (b) we find that when the ion coupling strength is relatively small (e.g. $\Gamma_i=125$ in the liquid state), the long wavelength kinetic modes exhibit instability against an oblique modulational perturbation   in strongly coupled plasmas with weakly relativistic degenerate electrons (e.g. in an outer mantle of white dwarfs). However, as   $\Gamma_i$ increases (e.g. $\Gamma_i=202$ in the crystallized state), the instability domain for the wave number expands, and the system can exhibit oblique modulational instability for a wide range of values of $k$. From Figs. 3(a) and (b) we can also conclude that the plane waves during propagation  in plasmas, e.g., in an outer mantle of white dwarf, exhibit stability against small perturbations and for a wide range of values of   $\alpha$ (where $\alpha$ is the angle, the wave vector $\mathbf{k}$ makes with the $x$-axis) as well as the wavelength. }  
  {Furthermore, when the modulation takes place along the direction of propagation of waves (i.e. for $\alpha=0$) in strongly coupled plasmas with weakly or ultra-relativistic degenerate electrons, the system always shows stability regardless of the values of $\Gamma_i$. From Figs. 3(c) and (d) we find that the instability of plane waves   in strongly coupled plasmas with ultra-relativistic degenerate electrons (e.g. in the interior of white dwarfs)},   {depends strongly on the scale length}     {of excitation of the wave modes, the angle of modulation, the ion coupling parameter as well as the degeneracy pressure of electrons.   Thus,   strong coupling of ions ($\Gamma_i>100$) as well as the   ultra-relativistic degeneracy pressure of electrons favor the instability of plane waves against the oblique modulation. }

The instability rate $\Gamma_{\text{inst}}$ is calculated and the developments of the instability growth or decay are shown in Fig. 4.  {We find that the maximum value of the instability rate typically depends on the cubic as well as the nonlocal nonlinearity through the coefficients $Q_1$ and $Q_2$ respectively. It is clear that the electron degeneracy has an important effect for the change of sign of the instability rate $\Gamma_{\text{inst}}$. For the waves propagating in   strongly coupled plasmas with weakly relativistic degenerate electrons, the plane wave perturbation grows exponentially with the wave number leading to the instability growth as in Fig. 4 (upper panel). However, a lower value of $\Gamma_i$ (when ions are in the liquid state)   tends to suppress the instability, decreasing the growth rate at lower value of the wave number $k$. This means that the plane waves in plasmas with crystallized ions and weakly relativistic degenerate electrons exhibit long wave MI. On the other hand, for the propagation of waves in strongly coupled plasmas with ultra-relativistic degenerate electrons, the instability rate becomes negative by the electron degeneracy. In this case, the contribution from the nonlocal nonlinearity becomes larger than the cubic nonlinear term leading to exponential decay of the perturbations, and the decay rate can not be suppressed by the strong coupling of ions, i.e., whenever ions are in liquid or crystallized states.  Thus,  while the growth rate  can be suppressed (i.e. $\Gamma_{\text{inst}}<1$) by lowering  the coupling parameter $\Gamma_i$ in the range $100<\Gamma_i\lesssim200$ (in the weakly relativistic case, see the upper panel of Fig. 4) with a cut-off at $K<1$, the decay rate remains unbounded within $K<1$, and can not be controlled  with a considerable range of values of $\Gamma_i$.  As an example, the dotted line in the lower panel of Fig. 4 has the cut-off at $K=7.4$ for $\Gamma_i=512$ corresponding to $Z_i=14$. The other lines (solid and dashed) corresponding to $\Gamma_i=125$, $202$ have cut-offs at higher $K>7.4$.}     
\section{Evolution of wave packets}
In order to examine a long-time evolution of the wave packets, we perform a numerical simulation of Eqs. \eqref{NLSE-1} and \eqref{NLSE-2} in the weakly relativistic regimes. To this end, we discretize the derivatives using a finite difference scheme, and use the domain size  {$25\leq\xi,\eta\leq25$}  with $150\times150$ grid points and time step $d\tau=10^{-3}$ for (2+1) dimensional evolution. In the case of (3+1)-dimensional evolution, we use  the domain size  {$15\leq\xi,\eta,\zeta\leq15$} with $60\times60\times60$ grid points and time step $d\tau=10^{-3}$. A symmetric Gaussian  wave beam is chosen as an initial condition of the form $\Phi_{\gamma}=\sqrt{2I/\pi ab}\exp(-\xi^2/a^2-\eta^2/b^2)$ in (2+1) dimensions [similar for (3+1) dimensions], where $I$ is the wave action. Notwithstanding, we have made simulations with other initial beam profiles,   {however, the qualitative results remain almost the same}.  Similar analysis can also be done in the case of ultra-relativistic limit $(\gamma=4/3)$, however, one has to be careful about the coefficients which become larger in magnitude, and one might have to rescale those in order to perform numerical analysis with smaller step size for  the space and/or time.

\subsection{(2+1)-dimensional evolution}
 \begin{figure*}
\includegraphics[width=6in,height=4in,trim=0.0in 0in 0in 0in]{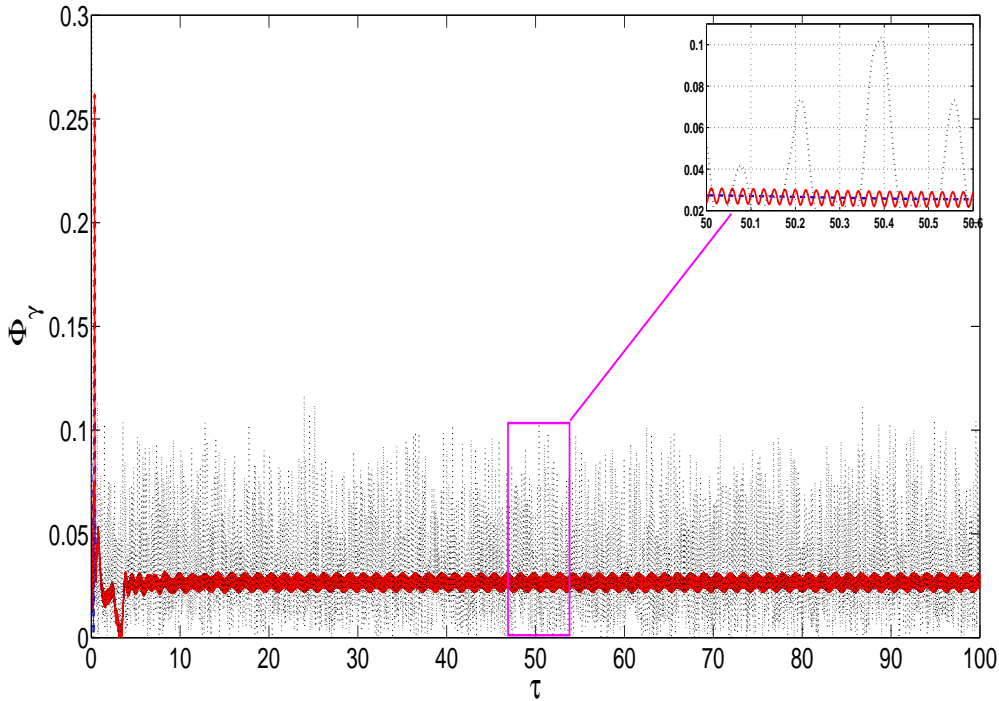}  
 \caption{ (Color online) Time evolution of [numerical solution of Eqs. \eqref{NLSE-1} and \eqref{NLSE-2}] the wave amplitude $\Phi_{\gamma}$ $(\gamma=5/3)$ for different $\Gamma_i$. The solid (red) and dashed (blue) (see the inset) lines represent stable oscillations, and correspond to $\bar{\nu}_{in}=0.1$ and  $\bar{\nu}_{in}=0$ respectively (with $k=0.5$, $\alpha=2$, $\Gamma_i=125$). The dotted line shows irregular oscillations for a different set of parameters, i.e., $\bar{\nu}_{in}=0$, $k=0.2$, $\alpha=1.6$ and $\Gamma_i=202$.}
\end{figure*}
\begin{figure*}
\includegraphics[width=6in,height=4in,trim=0.0in 0in 0in 0in]{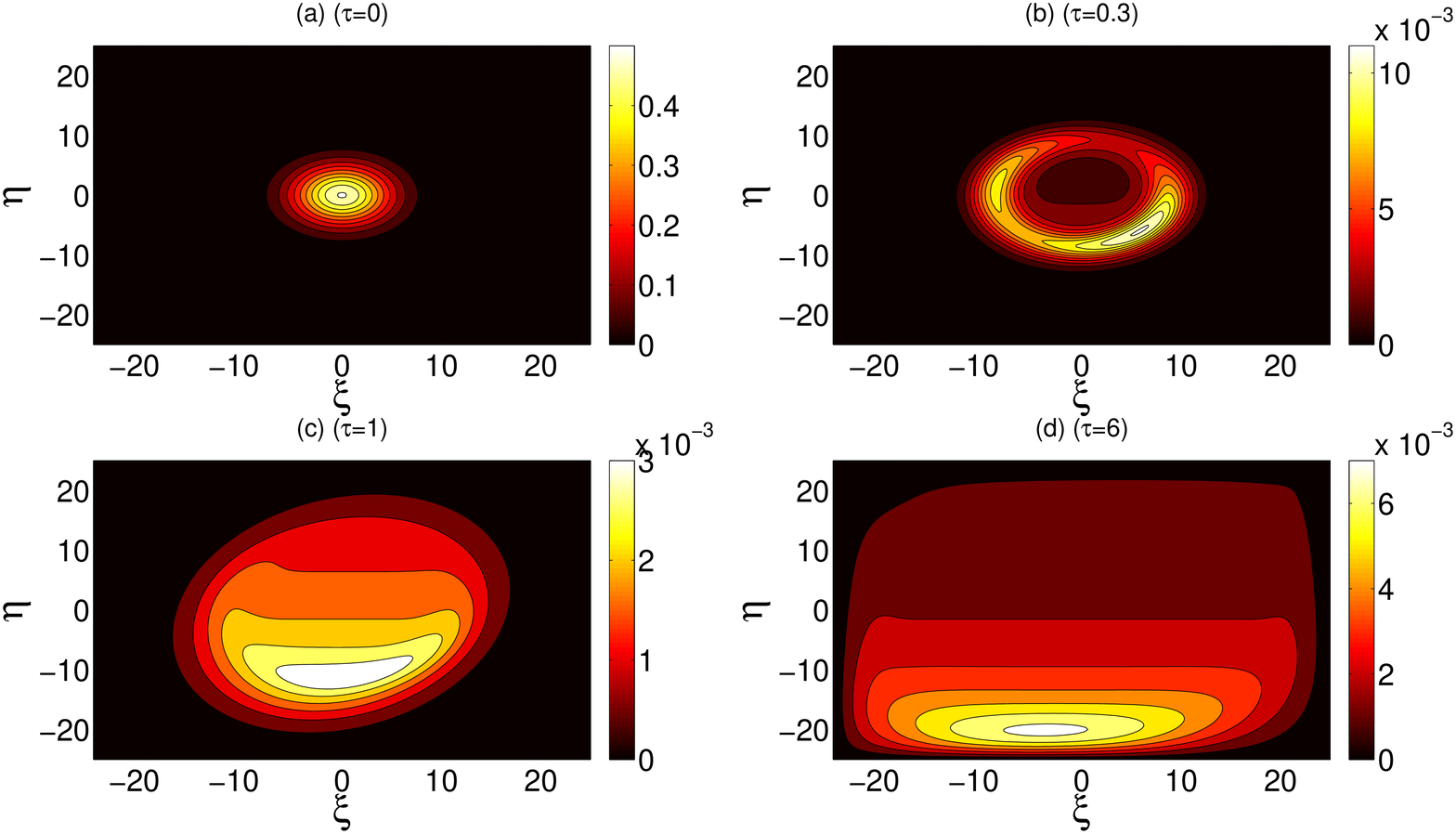}  
 \caption{ (Color online) Contour plots of a numerical solution of Eqs. \eqref{NLSE-1} and \eqref{NLSE-2} at different times as in the figure. The initial wave form decays due to dispersion until it gets modulated. The amplitude then grows until it takes the maximum modulation, and then propagates with a permanent profile. The parameter values are the same as for the stable solution, i.e., the solid line in Fig. 5.}
\end{figure*}
We neglect the $z$-dependence of the physical quantities. Then the coefficients $P_3$, $R_3$ and $S_3$ are all zero, and   Eqs. \eqref{NLSE-1} and \eqref{NLSE-2} reduce to Davey-Stewartson-like equations in a more generalized form. The latter, in particular, has been investigated by a number of authors (see, e.g. some Refs. in \cite{mi-ur,MI-DS})  {not only in the context of  plasmas}, but also in some other nonlinear media. Such equations  {may eventually give rise to unstable solutions other than localization, or the formation of a singularity} at which the solution blows-up with higher amplitudes in a shorter scale. This means that system's validity breaks down near the singularity implying that an additional physical mechanism might be necessary in order to arrest such blow-up. Here we show that no singularity is formed for a wide range of parameters appropriate for the model, i.e., the model that we have considered is self-consistent in (2+1) dimensions, and does not give any unbounded solution. 

Figure 5 shows the time evolution of the wave amplitude with two different values of the coupling parameter $\Gamma_i=125$ (blue or dashed line) and $202$ (black dotted line) with (solid or red line) and without (dotted and dashed lines) collisional effects. The solid or red line represents the curve with weak collisional effects and with the same $\Gamma_i=125$ as the dashed line. A steady state oscillation (see the inset), (even when the wave action is well above the critical power) is found, to be called  driven damped oscillations. Usually, due to the the collisional term, which acts like a frictional force, the wave amplitude dies away. However, in this case, the nonlocal nonlinearity always feeds the energy into the system so as to offset the frictional losses. Since the collision frequency is   smaller than the ion plasma frequency and $\bar{\nu}_{in}\sim\epsilon$, i.e., $Q_3\sim\epsilon/2$, higher values of $\bar{\nu}_{in}$ are inadmissible, and we still have a stable solution.

For a different set of values of the coefficients $P_1$, $P_2$, etc. with a higher $\Gamma_i=202$, the wave amplitude oscillates in an irregular manner (see the dotted line in Fig. 5), implying that the perturbations are not stable, and the wave packets may not be localized. Because, in contrast to the case of the solid or dashed curves (in which $P_1$, $P_2$, $P_4\sim Q_1$ and $Q_2<Q_1$), the dispersion coefficients become larger than the nonlinear terms, and either the cubic or quadratic nonlinearity is not enough to balance the higher dispersive effects.  {Thus, the localization of wave packets may be possible  in plasmas with ions   in liquid state and degenerate electrons are weakly relativistic. However, for plasmas in which ions are in crystallized state and degenerate electrons weakly or ultra-relativistic, the wave packets may not be localized with higher values of $\Gamma_i\sim200$. This is expected as we have seen in the previous section that the MI growth rate of plane waves can be suppressed by lowering the coupling parameter $\Gamma_i$ instead of its higher values in the weakly relativistic strongly coupled regimes. The higher values of $\Gamma_i$ gives rise to the enhancement of the growth rate leading to an exponential growth of perturbation with no cut-offs at a finite $k<1$. }

 {Figure 6 shows  examples} of time evolution of wave packets in the case of stable oscillations of wave amplitude. We find that the wave  is localized and propagates with a permanent profile. The parameter values are considered as the same as for the solid line in Fig. 5. Initially,  the wave amplitude decays (and the symmetric Gaussian form breaks down) into different shape until the wave gets modulated. The wave amplitude then starts growing until the maximum modulation is achieved. After some time, the wave  gets stabilized and propagates with a permanent profile due to nice balance of the dispersion and nonlinearity.  {Thus, we conclude that the localization of wave packets  having kinetic ion modes as carrier modes in plasmas where strongly coupled ions are in liquid state and degenerate electrons are weakly relativistic such as those in an outer mantle of white dwarfs, is possible through the MI. These packets propagate in such plasmas with a permanent profile for a long time. On the other hand, plasmas with crystallized ions and relativistically degenerate electrons, such as those in the core of white dwarfs, can not support the propagation of such wave packets with a permanent profile.}
\subsection{(3+1)-dimensional evolution}
\begin{figure*}
\includegraphics[width=6in,height=4in,trim=0.0in 0in 0in 0in]{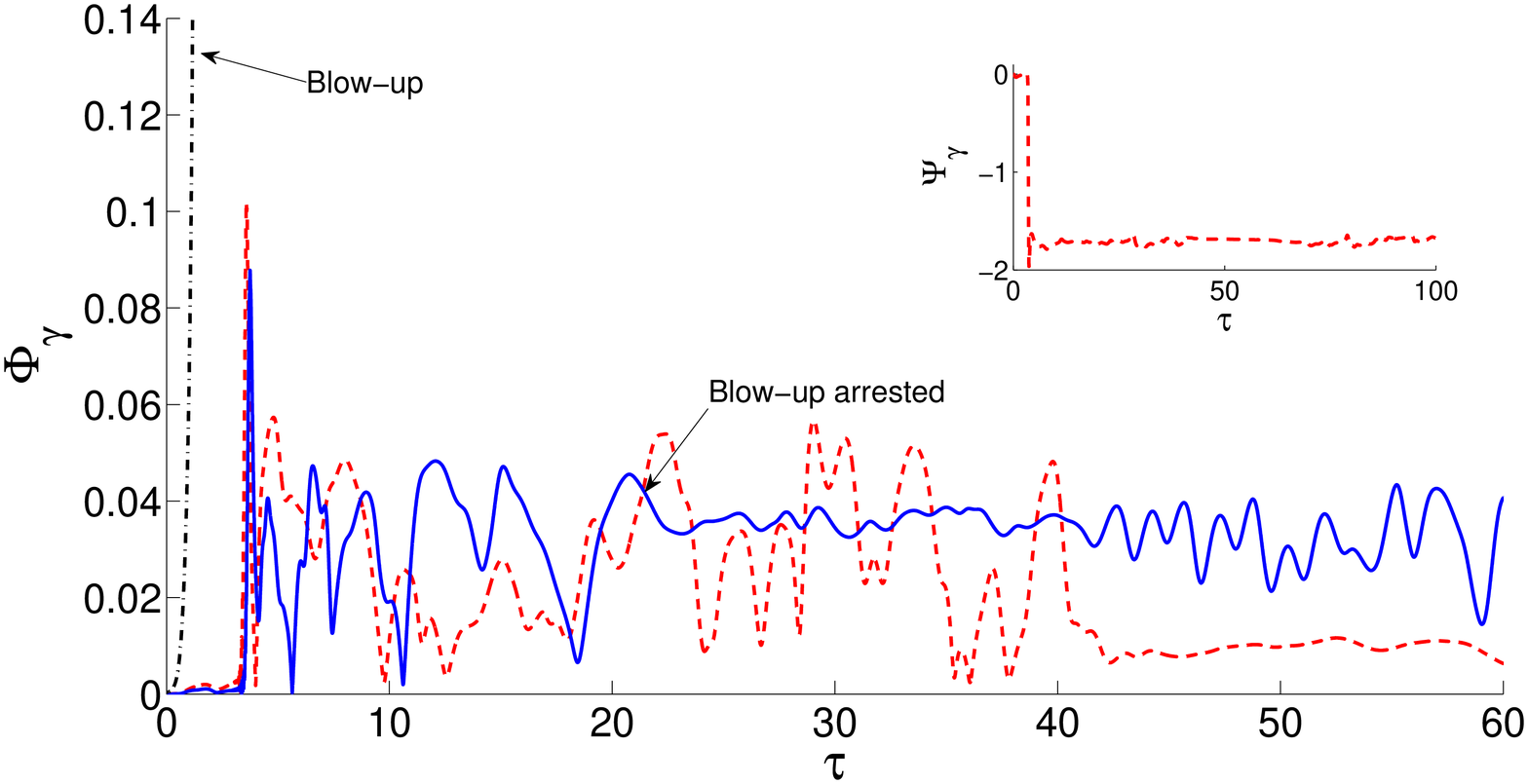}  
 \caption{ (Color online) Time evolution of [numerical solution of Eqs. \eqref{NLSE-1} and \eqref{NLSE-2}] the wave amplitude $\Phi_{\gamma}$ $(\gamma=5/3)$ for $\bar{\nu}_{in}=0$. The dash-dotted  line shows that the solution blows-up in a finite time corresponding to the same parameters, i.e., $k=0.5$, $\alpha=2$ and $\Gamma_i=125$  as for the dashed line (stable solution) in Fig. 5. The solid (blue) and dashed (red) lines show that the collapse is arrested with unstable or irregular oscillations (bounded) corresponding to $k=0.2$, $\alpha=1.6$, $\Gamma_i=202$ (same parameters as for the dotted line in Fig. 5), and a different set, i.e., $k=0.2$, $\alpha=2$, $\Gamma_i=125$ respectively. The inset shows  a profile for $\Psi_{\gamma}$ corresponding to the dashed (red) line for $\Phi_{\gamma}$. No stable solution is found.}
\end{figure*}
We  examine the effects of additional dispersive and nonlinear effects that are due to the $z$-dependence of the physical variables. We find that a singularity is formed at which the wave amplitude blows up in a short time (see the dash-dotted line in Fig. 7) for the same set of parameters as  the case of  a steady state solution in (2+1)-dimension [see dashed (blue) line in the inset of Fig. 5]. In this case, the coefficients are $P_1=1.59$, $P_2=1.3$, $P_3=1.68$, $P_4=0.9$, $Q_1=-1.4$, $Q_2=0.35$,  $Q_3=0.0$,  $R_1=-1.8$, $R_2=-1.16$, $R_3=-1.97$, $R_4=-0.74$,  $S_1=0.75$, $S_2=-2.1$, 
$S_3=1.5$ and $S_4=-8.0$. That is, additional effects due to the dispersion ($P_3$, $R_3$) and nonlinearity ($S_3$) are responsible for the formation of singularity. It turns out that the validity of Eqs. \eqref{NLSE-1} and \eqref{NLSE-2} in (3+1) dimensions breaks down near this singular point. Since a physical quantity can not be infinite,  some of these dispersive and/or nonlinear effects, which were initially small, become important in order to prevent such  collapse.  

We find that when the dispersive effects are more pronounced than the nonlinearities, which occurs  for a different set of parameters, i.e. of $k$ and/or the obliqueness $\alpha$ or by increasing the coupling parameter $\Gamma_i$ [e.g. $k=0.2$, $\alpha=2$, $\Gamma_i=125$ (dashed line in Fig. 7) or $k=0.2$, $\alpha=1.6$, $\Gamma_i=202$ (solid line in Fig. 7)], the collapse is arrested  with an unstable oscillation.    We do not find any steady state solution for a wide range of parameters. The latter, however,  requires further investigation  for a conclusive evidence, and is limited to the present study. The weak ion-neutral collision has no  effect other   than a weak damping of oscillations (not shown in the figure) of the wave amplitude. 
\section{Discussion and conclusion}
Results of the previous sections demonstrate that plasmas with strongly coupled ions and weakly or ultra-relativistically degenerate electrons can support the excitation of low-frequency kinetic ion wave modes.  The latter  with short-wavelengths  travel with a higher or lower group velocity than those with longer wavelengths depending on whether the  electrons are weakly relativistic or ultra-relativistically degenerate. This anomalous group velocity dispersion is one of the most important conditions for the modulational instability of wave packets in nonlinear media. We have discussed the parameter regimes in two cases in which the present model is valid. These regimes are, in particular, representative of the interior (e.g., carbon-oxygen composition) or an outer mantle  of white dwarfs.

  {In the present model, the equilibrium of electrons has been considered }to be maintained by  two pressure equations as per Chandrasekhar \cite{compact-objects-Pressure}, whereas that of ions are associated with the strong coupling effects \cite{sc-pressure1,sc-pressure2}.   We find that the dominant contribution of the effective ion temperature is mainly due to the strong interaction of ions. In absence of the latter, the wave modes propagate with  wavelengths above a critical value, and its contribution in the wave  group dispersion is higher than the kinetic temperature of ions. Since the relaxation time has been found to increase highly with the ion coupling parameter $\Gamma_i>10$ \cite{sc-viscosity},    only kinetic wave modes $(\omega\tau_m\gg1)$ exist in plasmas where ions form a liquid state or a crystallized one. In the latter, the elasticity will dominate over the viscosity effects.  The hydrodynamic modes, on the other hand, may be relevant in other regimes where $\Gamma_i$ is not so large, e.g., in $1<\Gamma_i<100$. In this case, the medium behaves like a liquid where the viscosity effects become important, and the wave frequency $\omega$ has to be well below the ion plasma frequency in order to satisfy the hydrodynamic limit. Also, for lower values of $\omega$ where the dispersion is weak, the soliton (e.g. Korteweg-de Vries soliton) formation of the carrier waves is a  lower order process than the modulational instability of the wave envelopes discussed here. 

We have studied the modulation of a waveform with constant power by perturbing its amplitude with a plane wave disturbance in multi-dimensional form. It shows that the obliqueness of modulation in the $xy$ plane destabilizes the wave packet whereas it remains stable under the parallel modulation $(\alpha=0)$ in both the weakly relativistic and ultra-relativistic regimes. Furthermore, in contrast to (1+1)-dimension,  the    wave number of perturbations   in multi-dimensional propagation, changes the domain of stable and unstable wave numbers significantly. This implies that what is stable in (1+1)-dimension could become unstable in a more general situation.  For the parameters as in Case I, the wave packet is shown to be stable within a wide range of  the carrier wave number  $k$ and   the obliqueness parameter $\alpha$. The  region of instability may be increased in the $k\alpha$ plane with  higher values of $\Gamma_i$. In these higher values, the growth rate of instability remains positive and high, which, however, can be suppressed by   a lower value of $\Gamma_i$ with a lower cut-off   of the wave number of perturbation $K<1$. On the other hand, the results corresponding to Case II show that the wave envelope is modulational unstable for a wide range of $k$ and $\alpha$. In this case, though the decay rate can be lowered by increasing $\Gamma_i$, however, it remains outside the considerable range  of $K<1$. Thus, for the propagation of wave packets in plasmas  {with ions  in liquid or crystallized states} as in the core of white dwarfs, the  decay rate of instability  may not be controlled. However, in plasmas with liquid state of ions and non-relativistic degenerate electrons (as in an outer mantle of white dwarfs), the waves are mostly stable to modulational perturbation, and the instability can even be suppressed with a lower value of $\Gamma_i$.   {This implies that the localization of wave packets having kinetic ion modes as the carrier waves associated with the MI may be possible in plasmas (e.g. in an outer mantle of white dwarfs) in which ions are in liquid state and degenerate electrons are weakly relativistic. Since the growth or decay rate of instability can not be suppressed by higher values of $\Gamma_i$, the MI does not give rise to localization of wave envelopes in plasmas where ions are in crystallized states and electrons are relativistically  degenerate (e.g. in the core of white dwarfs).}

Next, {we examine whether the localization of wave packets is possible though the MI in some regimes where   the instability growth rate is suppressed, and in some other regimes in which the time evolution of the wave amplitude exhibit irregular oscillations leading to delocalization of the wave envelopes.} The time evolution of the (2+1)-dimensional wave packets shows that the wave amplitude stabilizes for a long time with a lower value of the coupling parameter $\Gamma_i\sim125$. However, for a higher value of  $\Gamma_i\sim202$, it shows irregular oscillations (see Fig. 5), i.e., unstable.  {In the case of stable wave oscillations, we find that for an initial Gaussian wave beam}, the amplitude initially decays and the beam looses its shape before it gets modulated by the nonlinear self-interactions. The amplitude then starts increasing until   the maximum modulation is reached. As the time progresses, the wave amplitude reaches a steady state value, and the initial beam    {transforms into  another} with a permanent profile (see Fig.6). We have included a phenomenological ion-neutral collision term which appears in the third-order corrections as a linear term with the first order perturbation, and remains weaker than the nonlinear (Kerr and nonlocal) contributions. It has no effect on the stability or instability of wave modes under modulation, rather it changes the   oscillation pattern with a higher amplitude like a steady state damped harmonic oscillation (see solid line in Fig. 5).  {The frictional force due to  the collision, which usually damps the wave amplitude, does not make so   as  the nonlocal nonlinearity feeds up the sufficient energy} in order to offset  the decay.     

On the other hand, the results in the (3+1)-dimensional evolution indicate that a wave singularity can be formed leading to wave collapse in a finite time when the dispersive effects are not in balance with the nonlinearities, or the additional dispersion (due to $z$ coordinate) dominates over the group velocity dispersion. This collapse can, however, be arrested for a different choice of $k$ and/or $\alpha$, or by increasing the coupling strength $\Gamma_i$ where the dispersion is more pronounced than the nonlinearities. The numerical simulation reveals that   {the present model in (3+1) dimensions does not support the formation of localized structure} with stable oscillations. In order to obtain such steady state solution, one way could be to consider $v_g$ in an arbitrary direction of space or along a fixed axis.  However, one needs further investigation in order to confirm it, which is limited to the present study.   

To conclude,  in  strongly coupled plasmas with relativistically degenerate electrons, one should carefully consider the parameter regimes as discussed in  Case I and Case II. In  high coupling regimes $(\Gamma_i\gg1)$,  the effective temperature  associated with the strong electrostatic interactions of ions is much more pronounced than the kinetic ($T_i$) one. The modulation of plane waves and their nonlinear evolution depend strongly on the system parameters including the obliqueness $\alpha$, the Coulomb coupling $\Gamma_i$ as well as  the extra dimension (geometry) of the system.  {We find that the localization of wave packets in (2+1)-dimensions having kinetic ion modes as carrier waves is possible   in plasmas with ions in liquid state and degenerate electrons are weakly relativistic. In other plasma regimes, e.g., in the core of white dwarfs where ions are in crystallized state and degenerate electrons are ultra-relativistic, the MI of plane waves gives rise an unbounded growth or decay of instability leading to delocalization of wave envelopes.} Since $\Gamma_i$ can be controlled by heating or cooling the ion components, it would be interesting to look for the excitation of ion wave modes, and their localization as wave packets in laboratory experiments. Our model is more general than those available in the literature \cite{solitary-UR2,MI-sc2}, and could  be  applicable to other nonideal systems, e.g. metal plasmas.   
\section*{acknowledgments}  APM is thankful to the Kempe Foundations, Sweden for support.
\section*{Appendix A: Second order first harmonic modes}
 For $n=2$, $l=1$ we have
 \begin{equation}
N^{(2)}_{e1}=A_{\gamma}\phi^{(2)}_{1}, \hskip5pt W^{(2)}_{1}=c_{12}\frac{\partial \phi^{(1)}_1}{\partial\zeta}\label{ne2-l1}
 \end{equation}
\begin{equation}
 N^{(2)}_{i1}=K^2_{\gamma}\phi^{(2)}_{1}-i2D_{\gamma}\left(k_x\frac{\partial \phi^{(1)}_1}{\partial\xi}+k_y\frac{\partial \phi^{(1)}_1}{\partial\eta} \right), \label{ni2-l1}
\end{equation}
\begin{equation}
\alpha\left(U^{(2)}_{1},V^{(2)}_{1}\right)=\alpha_{u,v}\phi^{(2)}_{1}+\alpha_{u\xi,v\xi}\frac{\partial \phi^{(1)}_1}{\partial\xi}+ \alpha_{u\eta,v\eta}\frac{\partial \phi^{(1)}_1}{\partial\eta}, \label{uv2-l1}
\end{equation}
where \begin{equation}
\alpha=B^2_{12}-A_{12x}A_{12y},\hskip5pt B_{12}=\frac{\left(\bar{\zeta}+\bar{\eta}/3\right)k_x k_y}{1-i\omega\bar{\tau}_m}, 
\end{equation}
\begin{equation}
A_{12(x,y)}=-i\omega+\frac{\bar{\eta}k^2+\left(\bar{\zeta}+\bar{\eta}/3\right)k^2_{(x,y)}}{1-i\omega\bar{\tau}_m},
\end{equation}
\begin{equation}
\alpha_{(u,v)}=C_{12(x,y)}A_{12(y,x)}-C_{12(y,x)}B_{12},
\end{equation}
\begin{equation}
\left(\alpha_{u\xi},\alpha_{u\eta}\right)=B_{12}(E_{12y},D_{12y})-A_{12y}(D_{12x},E_{12x}),
\end{equation}
\begin{equation}
\left(\alpha_{v\xi},\alpha_{v\eta}\right)=B_{12}(D_{12x},E_{12x})-A_{12x}(E_{12y},D_{12y}),
\end{equation}
\begin{equation}
C_{12(x,y)}=ik_{(x,y)}\left(D_{\gamma}+K^2_{\gamma}\right),
\end{equation}
\begin{equation}
C_{12}=-\frac{i\left(D_{\gamma}+K^2_{\gamma}\right)+{\omega K^2_{\gamma}\left(\bar{\zeta}+\bar{\eta}/3\right)}/{(1-i\omega\bar{\tau}_m)}}{\omega+i\bar{\eta}k^2/(1-i\omega\bar{\tau}_m)},
\end{equation}
\begin{widetext}
\begin{eqnarray}
&&D_{12(x,y)}=\frac{\omega K^2_{\gamma}k_{(x,y)}}{k^2(1-i\omega\bar{\tau}_m)}\left[(1-i2\omega\bar{\tau}_m)v_{g(x,y)}+i2\eta^{*}k_{(x,y)}\right]-\frac{D_{\gamma}+K^2_{\gamma}}{1-i\omega\bar{\tau}_m}\left[1-i\bar{\tau}_m\left(\omega+k_{(x,y)}v_{g(x,y)}\right)\right] \notag \\ 
&&\hskip10pt+i\frac{\left(\bar{\zeta}+\bar{\eta}/3\right) K^2_{\gamma}\omega k^2_{(y,x)}}{k^2(1-i\omega\bar{\tau}_m)}-2D_{\gamma}k^2_{(x,y)},
\end{eqnarray}
\begin{eqnarray}
&&E_{12(x,y)}=\frac{\omega K^2_{\gamma}k_{(x,y)}}{k^2(1-i\omega\bar{\tau}_m)}\left[(1-i2\omega\bar{\tau}_m)v_{g(y,x)}+i2\bar{\eta}k_{(y,x)}\right]+i\frac{\left(D_{\gamma}+K^2_{\gamma}\right)\bar{\tau}_mk_{(x,y)}v_{g(y,x)}}{1-i\omega\bar{\tau}_m} \notag \\ 
&&\hskip10pt+i\frac{\left(\bar{\zeta}+\bar{\eta}/3\right) K^2_{\gamma}\omega k_xk_y}{k^2(1-i\omega\bar{\tau}_m)}-2D_{\gamma}k_xk_y,
\end{eqnarray}
\end{widetext}
\section*{Appendix B: Zeroth harmonic modes}
 For  $l=0$ we obtain the zeroth harmonic modes from the coefficients of $\epsilon^2$ and $\epsilon^3$ in terms of $\phi^{(2)}_0$ and $\lvert\phi^{(1)}_1 \rvert^2$.
 \begin{equation}
 N^{(2)}_{e0}=N^{(2)}_{i0}=A_{\gamma}\phi^{(2)}_0+2B_{\gamma}\lvert\phi^{(1)}_1 \rvert^2, \label{ne0}
 \end{equation}
 \begin{eqnarray}
 &&\Delta N^{(2)}_{i0}=\frac{\partial U^{(2)}_{0} }{\partial \xi}+\frac{\partial V^{(2)}_{0} }{\partial \eta}+\frac{\partial  W^{(2)}_{0} }{\partial \zeta}\notag\\
 &&\hskip10pt+\frac{2\omega K^4_{\gamma}}{k^2}\left(k_x\frac{\partial}{\partial\xi}+k_y\frac{\partial}{\partial\eta} \right)\lvert\phi^{(1)}_1 \rvert^2, \label{ni0}
  \end{eqnarray}
 \begin{equation}
 \Delta U^{(2)}_{0}=(A_{\gamma}+D_{\gamma})\frac{\partial \phi^{(2)}_0}{\partial \xi}+B_x\frac{\partial\lvert\phi^{(1)}_1 \rvert^2}{\partial\xi}+C_x\frac{\partial\lvert\phi^{(1)}_1 \rvert^2}{\partial\eta}, \label{u0}
 \end{equation}
 \begin{equation}
 \Delta V^{(2)}_{0}=(A_{\gamma}+D_{\gamma})\frac{\partial \phi^{(2)}_0}{\partial \eta}+C_y\frac{\partial\lvert\phi^{(1)}_1 \rvert^2}{\partial\xi}+B_y\frac{\partial\lvert\phi^{(1)}_1 \rvert^2}{\partial\eta}, \label{v0}
 \end{equation}
 \begin{equation}
 \Delta W^{(2)}_{0}=(A_{\gamma}+D_{\gamma})\frac{\partial \phi^{(2)}_0}{\partial \zeta}+D_{\gamma}K^2_{\gamma}\frac{\partial\lvert\phi^{(1)}_1 \rvert^2}{\partial\zeta}, \label{w0}
 \end{equation}
where $\Delta\equiv v_{gx}\partial_{\xi}+v_{gy}\partial_{\eta}$ and $B_{x,y}$, $C_{x,y}$ are given by 
\begin{equation}
B_{x,y}=2B_{\gamma}+D_{\gamma}K^2_{\gamma}-\frac{\omega K^4_{\gamma}}{k^2}k_{x,y}\left(v_{g(x,y)}-\frac{\omega k_{x,y}}{k^2}\right),
\end{equation}
\begin{equation}
C_{x,y}=\frac{\omega K^4_{\gamma}}{k^2}k_{x,y}\left(v_{g(y,x)}-\frac{\omega k_{y,x}}{k^2}\right).
\end{equation}
We use the operator $\Delta$ once in Eq. \eqref{ni0}, and eliminate $N^{(2)}_{i0}$, $U^{(2)}_{0}$, $V^{(2)}_{0}$, $W^{(2)}_{0}$ by using Eqs. \eqref{ne0} and \eqref{u0}-\eqref{w0} to obtain  Eq. \eqref{nonlocal}.

\section*{Appendix C: Coefficients  of second order second harmonic modes}
The coefficients of the second order harmonic modes for $n=2,l=2$ are given as follows:
\begin{equation}
\theta_{x,y}=\omega+\frac{2i\left[\bar{\eta}k^2+\left(\bar{\zeta}+\bar{\eta}/3\right)k^2_{x,y}\right]}{1-i2\omega\bar{\tau}_m},
\end{equation}
\begin{equation}
\theta_{xy}=\frac{4i\left(\bar{\zeta}+\bar{\eta}/3\right)k_xk_y}{1-i2\omega\bar{\tau}_m},
\end{equation}
\begin{equation}
\theta_{1,2}=-K^2_{\gamma}k_{x,y}\left[D_{\gamma}+\frac{i\omega\bar{\tau}_m}{1-i2\omega\bar{\tau}_m}\left(D_{\gamma}+K^2_{\gamma}-\frac{\omega^2K^2_{\gamma}}{k^2}\right)\right],
\end{equation}
\begin{widetext}
\begin{equation}
A_{22}=\frac{1}{\theta_x}\left[2k_xB_{\gamma}-\theta_1-B_{22}\theta_{xy}+2k_xD_{22}\left(A_{\gamma}+D_{\gamma}(1+4k^2)\right)\right],\hskip5pt B_{22}=\frac{1}{\vartheta}(\vartheta_1-\vartheta_2),
\end{equation}
\begin{equation}
D_{22}=\frac{1}{\vartheta}\left[\omega k_x(B_{\gamma}-K^4_{\gamma})(\theta^2_{xy}-\theta_x\theta_y)-(\theta_1-2k_xB_{\gamma})(2k^2_x\theta_y-k_xk_y\theta_{xy})+(\theta_2-2k_yB_{\gamma})(k^2_x\theta_{xy}-2k_xk_y\theta_{x})\right],
\end{equation}
\begin{equation}
\vartheta=(A_{\gamma}+4k^2D_{\gamma})\left[4k^2_x(k_y\theta_{xy}-k_x\theta_y)-\omega k_x(\theta^2_{xy}-4\theta_x\theta_y)-2k_xk^2_y\theta_x\right]-4D_{\gamma}k_x(k^2_x\theta_y+k^2_y\theta_x-k_xk_y\theta_{xy}),
\end{equation}
\begin{equation}
\vartheta_1=2\left[(A_{\gamma}+4k^2D_{\gamma})(k^2_x-\omega\theta_x)+D_{\gamma}k^2_x\right]\left[k_x(\theta_2-2B_{\gamma}k_y)+2\omega\theta_{xy}(B_{\gamma}-K^4_{\gamma}) \right],
\end{equation}
\begin{equation}
\vartheta_2=\left[(A_{\gamma}+4k^2D_{\gamma})(2k_xk_y-\omega\theta_{xy})+2D_{\gamma}k_xk_y\right]\left[k_x(\theta_1-2B_{\gamma}k_x)+2\omega\theta_{x}(B_{\gamma}-K^4_{\gamma}) \right],
\end{equation}
\end{widetext}
\section*{Appendix D: Coefficients of Eq. \eqref{NLSE-1}}
Here we give the coefficients $P_1$, $P_2$ etc. as follows:
\begin{widetext}
\begin{equation}
P_{1,2}\equiv\frac{1}{2}\frac{\partial^2\omega}{\partial k^2_{x,y}}=\frac{1}{\tilde{\omega}}\left(\frac{\omega^2}{k^2}-\frac{D^2_{\gamma}k^2}{K^4_{\gamma}}-4A_{\gamma}D^2_{\gamma}\frac{k^2_{x,y}}{K^6_{\gamma}}-v^2_{g(x,y)}\left(1-\frac{\bar{\tau}_m\eta^{*}k^2v_{g(x,y)}}{(1-i\omega\bar{\tau}_m)^3} \right)-\frac{i2\eta^{*}k_{x,y}v_{g(x,y)}}{(1-i\omega\bar{\tau}_m)^2} \right),  
\end{equation}
\begin{equation}
P_3=\frac{1}{\tilde{\omega}}\left(\frac{\omega^2}{k^2}-\frac{D^2_{\gamma}k^2}{K^4_{\gamma}}\right),
\end{equation}
\begin{equation}
P_4=\frac{W_n}{K^2_{\gamma}\tilde{\omega}}\left(\omega+\frac{i\eta^{*}k^2}{1-i\omega\bar{\tau}_m}\right)-\frac{i}{\tilde{\omega}K^2_{\gamma}}\left(\omega+\frac{i\bar{\eta}k^2}{1-i\omega\bar{\tau}_m}\right)^{-1}\left(k_xA_{12y}Q_x+k_yA_{12x}Q_y-\frac{(\bar{\zeta}+\bar{\eta}/3)k_xk_y}{1-i\omega\bar{\tau}_m} (k_yQ_x+k_xQ_y) \right),
\end{equation}
\begin{eqnarray}
&&Q_1=-\frac{1}{\tilde{\omega}K^2_{\gamma}}\left[R_{xy}\left(\omega+\frac{i\eta^* k^2}{1-i\omega\bar{\tau}_m}\right)+k^2\left(3C_{\gamma}-2B_{\gamma}D_{22}\right)+2B_{\gamma}\left(D_{\gamma}k^2-\omega^2 K^2_{\gamma}\right) \right. \notag\\
&&\left. +i\left(\omega+\frac{i\bar{\eta} k^2}{1-i\omega\bar{\tau}_m}\right)^{-1} \left(k_xR_xA_{12y}+k_yR_yA_{12x}-\frac{\left(\bar{\zeta}+\bar{\eta}/3\right)k_xk_y}{1-i\omega\bar{\tau}_m}\left( k_yR_x+k_xR_y\right)\right) \right],
\end{eqnarray}
\begin{equation}
Q_2=\frac{1}{\tilde{\omega}K^4_{\gamma}}\left[k^2(D_{\gamma}+K^2_{\gamma})(2B_{\gamma}-\omega A_{\gamma}K^2_{\gamma})-k^2K^2_{\gamma}(2B_{\gamma}+A_{\gamma}D_{\gamma})+A_{\gamma}\omega^2 K^4_{\gamma} \right],\hskip5pt Q_3=\frac{\omega\bar{\nu}_{in}}{\tilde{\omega}},
\end{equation}
\begin{equation}
W_n=C_u+B_v-2D_{\gamma}(k_yv_{gx}+k_xv_{gy}),
\end{equation}
\begin{eqnarray}
&&(B_u,C_v)=\left(\omega+\frac{i\eta^{*}k^2}{1-i\omega\bar{\tau}_m}\right)^{-1}\left(2D_{\gamma}k^2_{x,y}+\left(D_{\gamma}+K^2_{\gamma}\right)\left(1-\frac{i\bar{\tau}_mk_{x,y}v_{g(x,y)}}{1-i\omega\bar{\tau}_m}\right) \right.
\notag \\
&& \left.-\frac{\omega K^2_{\gamma}k_{x,y}}{k^2(1-i\omega\bar{\tau}_m)}\left((1-i2\omega\bar{\tau}_m)v_{g(x,y)}+i2\eta^{*}k_{x,y}\right)\right),
\end{eqnarray}
\begin{eqnarray}
&&(C_u,B_v)=\left(\omega+\frac{i\eta^{*}k^2}{1-i\omega\bar{\tau}_m}\right)^{-1}\left(2D_{\gamma}k_{x}k_y-\left( D_{\gamma}+K^2_{\gamma}\right)\frac{i\bar{\tau}_m k_{x,y}v_{g(x,y)}}{1-i\omega\bar{\tau}_m}\right. \notag \\
&&\left.-\frac{\omega K^2_{\gamma}}{k^2(1-i\omega\bar{\tau}_m)}\left[k_{x,y}\left((1-i2\omega\bar{\tau}_m)v_{g(y,x)}+i2\bar{\eta}k_{y,x}\right)
+i2\left(\bar{\zeta}+\frac{\bar{\eta}}3\right)k_x k_y\right]\right),
\end{eqnarray}
\begin{equation}
Q_{x,y}=W_{x,y}+\frac{i}{1-i\omega\bar{\tau}_m}\left[\bar{\tau}_mv_{g(y,x)}\left(D_{\gamma}+K^2_{\gamma}-2\frac{\omega}{k^2}K^2_{\gamma}k_{x,y}v_{g(x,y)} \right)+\frac{\omega}{k^2}\left(\bar{\zeta}+\frac{\bar{\eta}}3\right)K^2_{\gamma}k_{y,x} \right],
\end{equation}
\begin{eqnarray}
&&W_{x,y}=\left(1-\frac{i\omega\bar{\tau}_m}{1-i\omega\bar{\tau}_m} \right)\left(C_{u,v}v_{gx}+B_{u,v}v_{gy}\right)+\frac{2i}{1-i\omega\bar{\tau}_m}\left[k_{x,y}(C_u,B_v)\eta^{*}+k_{y,x}(B_u,C_v)\bar{\eta} \right] \notag \\
&&+\frac{i\left(\bar{\zeta}+\bar{\eta}/3\right)}{1-i\omega\bar{\tau}_m}\left(k_xB_{v,u}+k_yC_{v,u}\right)-2D_{\gamma}\left(k_{y,x}-\frac{i\bar{\tau}_mk_{x,y}}{1-i\omega\bar{\tau}_m}(k_yv_{gx}+k_xv_{gy}) \right),
\end{eqnarray}
\begin{equation}
R_{xy}=\omega K^2_{\gamma}\left(2B_{\gamma}+C_{22}\frac{k^2_x}{k^2} \right)+K^2_{\gamma}(k_x A_{22}+k_yB_{22})-\omega\left(3C_{\gamma}+2B_{\gamma}D_{22}\right), \hskip5pt C_{22}=B_{\gamma}+\left(A_{\gamma}+4k^2D_{\gamma}\right)D_{22}, 
\end{equation}
\begin{eqnarray}
&&R_{x,y}=k_{x,y}\left[\frac{\omega K^2_{\gamma}}{k^2}\left(\omega\left(C_{22}+3K^4_{\gamma}\right)-k_xA_{22}-k_yB_{22}\right)+D_{\gamma}\left(2K^2_{\gamma}D_{22}-C_{22}\right)\right]\notag \\
&&-\frac{i\bar{\tau}_mK^2_{\gamma}}{1-i\omega\bar{\tau}_m}\left[2\omega\left[\omega\left(A_{22},B_{22}\right)-k_{x,y}\left(D_{\gamma}K^2_{\gamma}+2\left(D_{\gamma}D_{22}+C_{22}\right)\right)\right]+k_{x,y}\left(k_xA_{22}+k_yB_{22}\right)\left(1+\frac{D_{\gamma}}{K^2_{\gamma}}-\frac{\omega^2}{k^2}\right)   \right].
\end{eqnarray}
\end{widetext}  

\end{document}